# ZEBRA-Prop: A Zero-Shot Embedding-Based Rapid and Accessible Regression Model for Materials Properties

Ryoma Yamamoto[1*], Akira Takahashi[1*], Kei Terayama[2,3], Yu Kumagai[4], Fumiyasu Oba[1,3,5]

1. Materials and Structures Laboratory, Institute of Integrated Research, Institute of Science Tokyo, 4259 Nagatsuta, Midori-ku, Yokohama 226-8501, Japan
2. Graduate School of Medical Life Science, Yokohama City University, 1-7-29 Suehiro-cho, Tsurumi-ku, Yokohama 230-0045, Japan
3. MDX Research Center for Element Strategy, Institute of Integrated Research, Institute of Science Tokyo, 4259 Nagatsuta, Midori-ku, Yokohama 226-8501, Japan
4. Institute for Materials Research, Tohoku University, 2-2-1 Katahira, Aoba-ku, Sendai 980-8577, Japan
5. Kanagawa Institute of Industrial Science and Technology (KISTEC), 705-1 Shimoimaizumi, Ebina 243-0435, Japan

*E-mail: yamamoto.r.ff57@m.isct.ac.jp (R. Yamamoto), takahashi.a.f9db@m.isct.ac.jp (A. Takahashi)

**Abstract**

Large language models (LLMs) exhibit substantial potential across diverse scientific disciplines, including materials science. A property prediction framework, ZEBRA-Prop (Zero-Shot Embedding-Based Rapid and Accessible Regression Model for Materials Properties), is presented here as an extension of LLM-Prop. In contrast to LLM-Prop, which requires task-specific fine-tuning of the LLM, ZEBRA-Prop eliminates fine-tuning, thereby reducing computational cost and enabling rapid model training. The framework employs MatTPUSciBERT, an LLM specialized for materials science, to enhance predictive capability. Multiple textual embeddings are incorporated through a learnable weighting mechanism, which alleviates the context-length constraints inherent in LLM-Prop and facilitates effective integration of diverse textual representations. Evaluation is conducted using two datasets: the TextEdge dataset (approximately 140,000 entries) and an in-house dataset (approximately 2,000 entries) derived from the Materials Project database, with physical properties obtained from first-principles calculations. The predictive performance of ZEBRA-Prop is close to that of LLM-Prop, while the training time is reduced by approximately 95%. The performance improvements are attributable to three principal factors: domain-specific LLM utilization, diversified textual descriptions, and systematic text preprocessing. ZEBRA-Prop constitutes a scalable and computationally efficient framework for materials property prediction and supports accelerated materials discovery, particularly under limited computational resources.

# 1. Introduction

The application of large language models (LLMs) in materials science has been steadily expanding. In the early stages, models such as Word2Vec[1] and BERT[2], which convert text into semantic vector representations, attracted considerable attention primarily because they could directly process textual information. This capability is well suited to the nature of scientific research, where knowledge is primarily communicated through literature. These models have been applied to a range of literature-mining tasks, including named entity recognition, relation extraction, and abstract or paper classification (e.g., Mat2Vec[3] and MatBERT[4]). Furthermore, they have been used to enable text-based crystal structure generation and similarity searches across large textual corpora (e.g., SciBERT[5], MatBERT, and MatSciBERT[6]).

More recently, encoder-type models based on architectures like BERT, which derive semantic embeddings from input text through bidirectional encoding, are increasingly used in materials science. Furthermore, generative language models such as GPT[7,8] and LLaMA[9], which autoregressively generate natural language, have gained prominence. Cross-modal alignment approaches have also emerged that explicitly link textual and structural representations of crystals[10–12]. These developments have broadened the scope of applications to crystal-structure generation, materials property prediction, materials discovery, and even experimental automation[13,14].

Since LLMs operate directly on natural-language descriptions rather than complex descriptors for crystal structures, LLM-based frameworks are accessible even to researchers who do not specialize in computational or data-driven methods. Moreover, LLM-based frameworks can process text containing both numerical and categorical descriptions without the need to unify dimensions or formats through mathematical transformations. This flexibility enables effective incorporation of domain knowledge into the prediction process.

Among the many machine-learning applications in materials science, property prediction remains one of the most prominent. Accurate prediction of materials properties is fundamental to understanding structure–property relationships and plays a crucial role in guiding the design of new materials[15–18]. Conventionally, many studies have tackled this task using feature descriptors[19–22] or crystal-graph representations derived from materials-science insights[23–25].

In recent years, prediction methods that leverage LLMs have been proposed, and further advances are expected as LLM research in computer science continues to evolve[26–28]. One notable example of an LLM-based property-prediction model is LLM-Prop[26]. Based on the T5 encoder architecture[29], LLM-Prop first converts textual descriptions of crystal structures into numerical embedding vectors using an LLM, which is fine-tuned during training to better capture the relationship between textual representations and material properties. The resulting embeddings are then passed through linear layers to estimate property values. LLM-Prop was evaluated on the TextEdge dataset, which contains 144,931 entries from the Materials Project database[30] with associated properties such as the band gap, unit-cell volume, and direct/indirect gap classification. Across these tasks, LLM-Prop achieved predictive accuracies comparable to current state-of-the-art crystal graph-based models[26].

However, several limitations remain. For example, fine-tuning an LLM still demands substantial computational resources. This poses a significant barrier for researchers who do not specialize in computational methods, as they typically lack access to high-performance computing infrastructure. In addition, the limited context length that current LLMs can handle remains a bottleneck when processing comprehensive structure descriptions that incorporate multiple types of information. The design of textual representations also needs improvement to support a broader range of material properties.

In this study, we introduce ZEBRA-Prop (Zero-Shot Embedding-Based Rapid and Accessible Regression Model for Materials Properties), an LLM-based property-prediction framework designed to be accessible to non-informatics researchers. The proposed model does not require fine-tuning, enabling rapid training and deployment. In addition, it alleviates the context-length limitations typically associated with LLMs by integrating multiple embeddings through a learnable weighting mechanism. Furthermore, prediction accuracy is improved through text preprocessing, the use of LLMs specialized for the materials science domain, and the integration of information from descriptions that capture multiple aspects of material characteristics. We evaluate the performance of ZEBRA-Prop on both in-house datasets and the publicly available subset of the TextEdge dataset (hereafter simply referred to as the TextEdge dataset; see Section 4.5.1 for details of the subset), which was introduced in the LLM-Prop study, and we share insights into text-preprocessing strategies and other design considerations that underpin its effectiveness.

## 2. Results and Discussion

In the proposed ZEBRA-Prop framework (**Figure 1**), the property prediction model is constructed through the following four stages: (1) the crystal structure is first expressed as a textual representation composed of several sentences, each describing the structure from a different perspective; (2) each sentence is then individually converted into an embedding vector using materials-science-specific LLMs; (3) these embeddings are then integrated through a learnable weighted integration; and (4) the resulting aggregated embedding is fed into a multilayer perceptron (MLP) to predict the target property value.

A key advantage of ZEBRA-Prop is that users can design their own textual descriptions based on domain knowledge, without being constrained by predefined descriptor sets. In this study, we employ descriptions derived from established matminer[31] descriptors to enable a fair comparison with a descriptor-based baseline (random forest[32]) under identical information conditions. In addition, we incorporate natural-language crystal-structure descriptions generated by Robocrystallographer[33], demonstrating the flexibility of the framework in integrating diverse textual inputs. Specifically, we use a total of twelve input sentences for ZEBRA-Prop. These sentences are constructed by combining (1) descriptions derived from the matminer code, where each sentence consists of the descriptor generated by an individual featurizer along with its corresponding label (referred to as matminer descriptions), and (2) sentences corresponding to the *mineral* and *components* parts of the descriptions generated by Robocrystallographer (referred to as robocrys. descriptions). As part of the text preprocessing, numerical values are integerized (i.e., scaled and rounded to integers) by scaling across the dataset (see Section 4.3 for details). MatTPUSciBERT[34] is employed as the backbone of the model to improve prediction accuracy by leveraging domain-specific knowledge in the absence of task-specific fine-tuning.

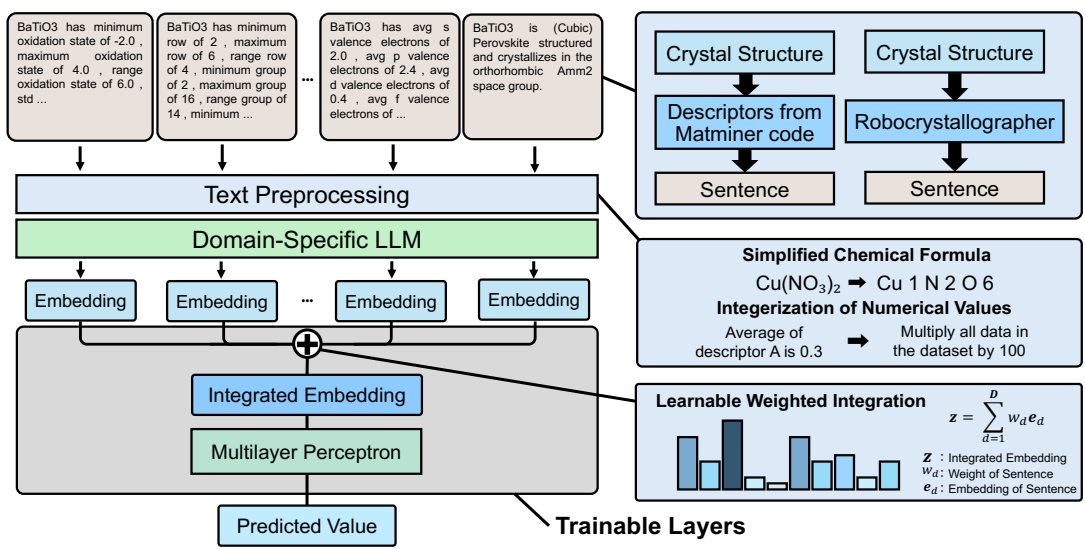

**Figure 1.** Schematic overview of ZEBRA-Prop.

## 2.1. Computational Efficiency and Prediction Accuracy

### 2.1.1. Training Time Comparison

While high predictive accuracy is a fundamental requirement for materials property prediction models, computational cost during training and inference is also an important consideration. In particular, LLM-based approaches such as LLM-Prop typically involve fine-tuning large-scale models with a large number of trainable parameters, which results in substantial training costs. High training costs can be prohibitive for researchers without access to large computational resources, and can hinder iterative model development for different datasets, input representations, and hyperparameters. To systematically compare models from the perspective of training cost, we first conducted training evaluations on an in-house dataset, which contains 2,202 material entries, using a single NVIDIA H100 GPU (see Section 4.5.3 for details). We then compared the training times of ZEBRA-Prop, LLM-Prop, graph neural network (GNN)-based models, including CGCNN (Crystal Graph Convolutional Neural Networks) and ALIGNN (Atomistic Line Graph Neural Network), and a decision-tree-based random forest model.

**Figure 2(a)** illustrates that ZEBRA-Prop achieves an approximately 95% reduction in training time compared to LLM-Prop and also requires less training time than CGCNN and ALIGNN. These results demonstrate the high computational efficiency of ZEBRA-Prop. On the other hand, in terms of training time alone, the random forest model is even more efficient than ZEBRA-Prop. However, ZEBRA-Prop can flexibly incorporate diverse textual descriptions, which leads to improved prediction accuracy as shown in Section 2.2.

This significant reduction in the training time of ZEBRA-Prop is achieved by avoiding fine-tuning of the LLM. Approximately 80% of the total training time is devoted to converting textual inputs into embedding representations using an LLM-based text encoder, while the remaining 20% is spent on training the weighting mechanism and the MLP regression model (**Figure 2(b)**). Importantly, the embedding vectors are computed only once and can be reused across different prediction tasks (e.g., different target properties or hyperparameters), which reduces the computational overhead of subsequent training. As a result, the effective retraining cost of ZEBRA-Prop is significantly lower than that of methods using LLM fine-tuning.

Due to this reduction in training time, ZEBRA-Prop does not rely on large-scale computational resources or specialized hardware such as high-performance GPUs. For a dataset of several thousand materials, the proposed framework remains computationally lightweight and can be executed on commodity hardware such as a standard laptop. We further confirmed the feasibility of running the training of ZEBRA-Prop on a laptop equipped with an Apple M2 chip and 16 GiB of unified memory (embedding conversion took several tens of minutes, and MLP training took a few minutes on the in-house dataset). Furthermore, API-based text embedding services could be used as an alternative to local LLM inference, further lowering the barrier to adoption across diverse hardware environments.

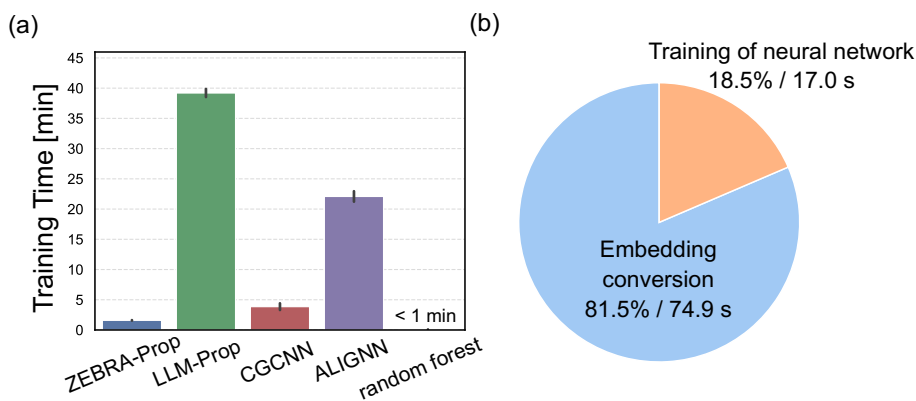

**Figure 2.** Computational efficiency of ZEBRA-Prop. (a) Comparison of training times for ZEBRA-Prop, LLM-Prop, CGCNN, ALIGNN, and a random forest. (b) Breakdown of the training time for ZEBRA-Prop. All values are averaged over 5-fold cross-validation on the in-house dataset. Error bars indicate ±1 standard deviation.

*2.1.2. Prediction Accuracy Comparison*

Next, we compared the predictive accuracy of each model using the in-house dataset and the TextEdge dataset. As shown in **Table 1**, ZEBRA-Prop achieves predictive performance close to that of LLM-Prop for the band gap, formation energy, and dielectric constants (both electronic and ionic contributions). By using both the descriptions generated from matminer descriptors and those generated from robocrys, the predictive accuracy of ZEBRA-Prop is improved, demonstrating its ability to effectively leverage multiple sources of information. This improvement clearly highlights an important advantage of LLM-based frameworks, namely their ability to directly process text containing both numerical and categorical descriptions without requiring unification of dimensions or formats through mathematical transformations. By encoding diverse descriptors in a textual form, ZEBRA-Prop can flexibly integrate information that is difficult to combine using conventional descriptor-based approaches.

Compared to LLM-based models, both CGCNN and ALIGNN outperform them, with ALIGNN achieving the highest predictive accuracy across all tasks. This performance gap can be attributed to the advantage of GNN-based approaches in explicitly incorporating detailed crystal structural information. Meanwhile, when ZEBRA-Prop is compared with the random forest, the predictive accuracies are nearly identical across all target properties. This result suggests that ZEBRA-Prop preserves most of the information contained in the original descriptors during the descriptor-to-text transformation process, or alternatively, that any potential information loss is compensated for by the knowledge implicitly acquired by the LLM during pretraining.

**Table 1.** Performance comparison of ZEBRA-Prop with baseline models on the in-house dataset. All values are averaged over the test folds in 5-fold cross-validation (see Section 4.5.4 for details). MAE and $R^2$ denote the mean absolute error and the coefficient of determination, respectively. The reported dielectric constants are relative dielectric constants, where "electronic" and "ionic" correspond to the electronic and ionic contributions. ZEBRA-Prop (w/o robocrys.) denotes the variant trained using only matminer descriptions without incorporating robocrys. descriptions. The parity plots for respective properties are shown in **Figure S2** (Supporting Information).

| Model | Band gap [eV] | | Formation Energy [eV / atom] | | Dielectric const. (electronic) | | Dielectric const. (ionic) | |
|---|---|---|---|---|---|---|---|---|
| | MAE↓ | $R^2$↑ | MAE↓ | $R^2$↑ | MAE↓ | $R^2$↑ | MAE↓ | $R^2$↑ |
| ALIGNN | **0.402±0.021** | **0.946±0.008** | **0.031±0.002** | **0.995±0.002** | **0.019±0.001** | **0.969±0.007** | **0.090±0.007** | **0.820±0.034** |
| CGCNN | 0.486±0.020 | 0.928±0.007 | 0.061±0.015 | 0.988±0.005 | 0.025±0.001 | 0.956±0.006 | 0.109±0.004 | 0.763±0.032 |
| random forest | 0.612±0.020 | 0.888±0.006 | 0.140±0.008 | 0.945±0.006 | 0.041±0.002 | 0.883±0.023 | 0.137±0.006 | 0.632±0.041 |
| LLM-Prop | 0.522±0.028 | 0.911±0.011 | 0.085±0.005 | 0.978±0.002 | 0.033±0.002 | 0.917±0.018 | 0.117±0.007 | 0.718±0.024 |
| ZEBRA-Prop (w/o robocrys.) | 0.599±0.025 | 0.882±0.011 | 0.103±0.003 | 0.966±0.003 | 0.042±0.003 | 0.880±0.013 | 0.138±0.007 | 0.635±0.025 |
| ZEBRA-Prop | 0.571±0.032 | 0.894±0.012 | 0.102±0.003 | 0.969±0.002 | 0.038±0.002 | 0.895±0.016 | 0.123±0.007 | 0.693±0.026 |

The comparison of prediction accuracy for band gap and formation energy on the TextEdge dataset is made in **Table 2**. LLM-Prop achieves predictive performance comparable to that of ALIGNN, which exhibits the highest accuracy among the evaluated models. The band gap results are consistent with the trends reported in the original LLM-Prop study. While ZEBRA-Prop does not reach the accuracy of ALIGNN or LLM-Prop, it achieves comparable or better accuracy than CGCNN and random forest. In addition, for formation energy prediction, which was newly incorporated into the TextEdge dataset in this study, ALIGNN demonstrates the best predictive performance as well. ZEBRA-Prop shows reasonable accuracy, though lower than that of CGCNN and LLM-Prop. These results, combined with the reduced training time discussed in Section 2.1.1, demonstrate that ZEBRA-Prop provides a reasonable balance between efficiency and accuracy.

**Table 2.** Performance comparison of ZEBRA-Prop with baseline models on the TextEdge dataset. All values are averaged over the test folds in 5-fold cross-validation. ZEBRA-Prop (w/o robocrys.) denotes the variant in which ZEBRA-Prop is trained using only matminer descriptions, without incorporating robocrys. descriptions. The parity plots for respective models and properties are shown in **Figure S3** (Supporting Information).

| Model | Band gap [eV] | | Formation Energy [eV / atom] | |
|---|---|---|---|---|
| | MAE↓ | $R^2$↑ | MAE↓ | $R^2$↑ |
| ALIGNN | **0.238±0.003** | **0.868±0.004** | **0.032±0.001** | **0.996±0.001** |
| CGCNN | 0.347±0.019 | 0.837±0.006 | 0.089±0.031 | 0.988±0.005 |
| random forest | 0.390±0.003 | 0.800±0.004 | 0.134±0.002 | 0.932±0.006 |
| LLM-Prop | 0.256±0.003 | 0.851±0.004 | 0.073±0.009 | 0.980±0.003 |
| ZEBRA-Prop (w/o robocrys.) | 0.345±0.004 | 0.784±0.004 | 0.123±0.002 | 0.939±0.007 |
| ZEBRA-Prop | 0.314±0.005 | 0.820±0.004 | 0.107±0.001 | 0.968±0.002 |

**2.2. Analysis of the Weight Integration Mechanism**

Next, to evaluate the effectiveness of the weighted integration mechanism in ZEBRA-Prop, we compare a model trained on all textual descriptions together with models trained on each description individually. As shown in **Figure 3** on the in-house dataset, incorporating all descriptions consistently achieves higher prediction accuracy than any single description setting. This result indicates that ZEBRA-Prop's weighted integration mechanism effectively aggregates complementary information across multiple textual perspectives.

Furthermore, the same evaluation on the TextEdge dataset demonstrates that the weighted integration mechanism significantly improves the accuracy of band gap prediction, as illustrated in **Figure 4**. This behavior is attributed to the characteristics of the dataset. In contrast to the in-house dataset, which contains only thermodynamically stable phases (i.e., those on the convex hull) and thus includes only one phase per composition, the TextEdge dataset includes unstable and metastable phases of the same composition. In such cases, composition-based descriptors (e.g., ElementProperty) alone are insufficient to capture property variations originating from differences in crystal structure. Explicitly incorporating information related to atomic configurations is therefore necessary. When the dataset encompasses structurally diverse phases, incorporating multiple complementary descriptors is often advantageous for improving predictive accuracy, whether the model is text-based or not. In addition, the relatively large standard deviation of the $R^2$ score when using only the *mineral* description is due to the presence of a fold with particularly large prediction errors (MAE: 0.952 eV, $R^2$: -0.128).

The relationship between single-sentence prediction performance and the learned integration weights is shown in **Figure S4** and **Figure S5** (Supporting Information). The learned weights are not necessarily proportional to the predictive accuracy of the corresponding individual sentences. In other words, sentences that exhibit higher standalone prediction performance do not always receive larger weights in the integrated model. One possible explanation for this discrepancy is semantic redundancy among the sentence descriptions. As shown in **Figure S6** (Supporting Information), several descriptions exhibit high embedding similarity, indicating that they encode overlapping information. When multiple sentences convey similar structural or compositional features, the model may distribute the weights across them rather than assigning a dominant weight to a single high-performing sentence. This suggests that the weighted integration mechanism does not simply prioritize individually accurate descriptions but instead balances complementary and redundant information to optimize overall prediction performance.

To further analyze why the weighted integration mechanism is effective, **Figure 5** presents a comparison of ZEBRA-Prop on formation energy prediction using the TextEdge dataset, contrasting the proposed weighted integration approach with models trained on each textual description individually.

As seen in **Figure 5(a)**, a distinct cluster of outliers is observed in the region where the ground-truth or predicted formation energy lies between −1 and 0 eV/atom. Based on this observation, we define EP-based outliers as data points in this region for which the prediction error exceeds 1.5 eV/atom when using only ElementProperty descriptions. Similarly, **Figure 5(b)** shows that, even after excluding EP-based outliers, there remain data points with substantial prediction errors when using only GlobalSymmetryFeatures descriptions. We define these as GSF-based outliers, namely data points not included in EP-based outliers for which the absolute prediction error exceeds 2.0 eV/atom.

The results show that when relying solely on ElementProperty descriptions (**Figure 5(a)**), large prediction errors persist for both EP-based outliers and GSF-based outliers. In contrast, descriptions incorporating structural information, such as GlobalSymmetryFeatures descriptions (**Figure 5(b)**), substantially improve prediction accuracy for the EP-based outliers. Furthermore, using *components* descriptions (**Figure 5(c)**), which encode more detailed structural information, leads to additional improvements, even for the GSF-based outliers, where predictions based on GlobalSymmetryFeatures descriptions remain inaccurate. Finally, when all textual descriptions are integrated through the weighted integration mechanism (**Figure 5(d)**), ZEBRA-Prop achieves consistently high prediction accuracy across both EP-based outliers and GSF-based outliers. Notably, the quantitative accuracy metrics are significantly improved relative to the results obtained using *components* descriptions alone. For completeness, the parity plots corresponding to the other descriptions are provided in **Figure S7** (Supporting Information), as they exhibit trends consistent with those discussed above.

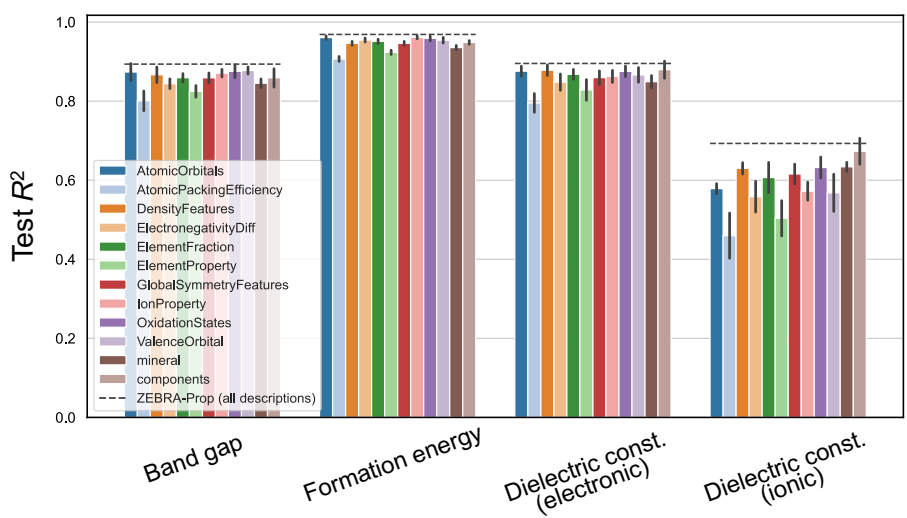

**Figure 3.** Coefficient of determination ($R^2$) on the test folds for ZEBRA-Prop trained with individual input descriptions (bar plot) and with all input descriptions combined (dashed line), for each property on the in-house dataset. Values are averaged over test folds in 5-fold cross-validation.

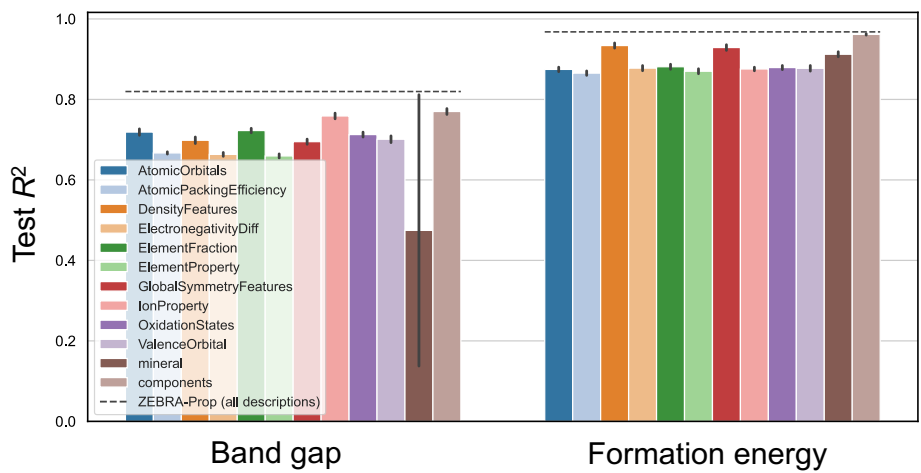

**Figure 4.** Coefficient of determination ($R^2$) on the test fold for ZEBRA-Prop trained with individual input descriptions (bar plot) and with all input descriptions combined (dashed line), for each property on the TextEdge dataset. Values are averaged over test folds in 5-fold cross-validation.

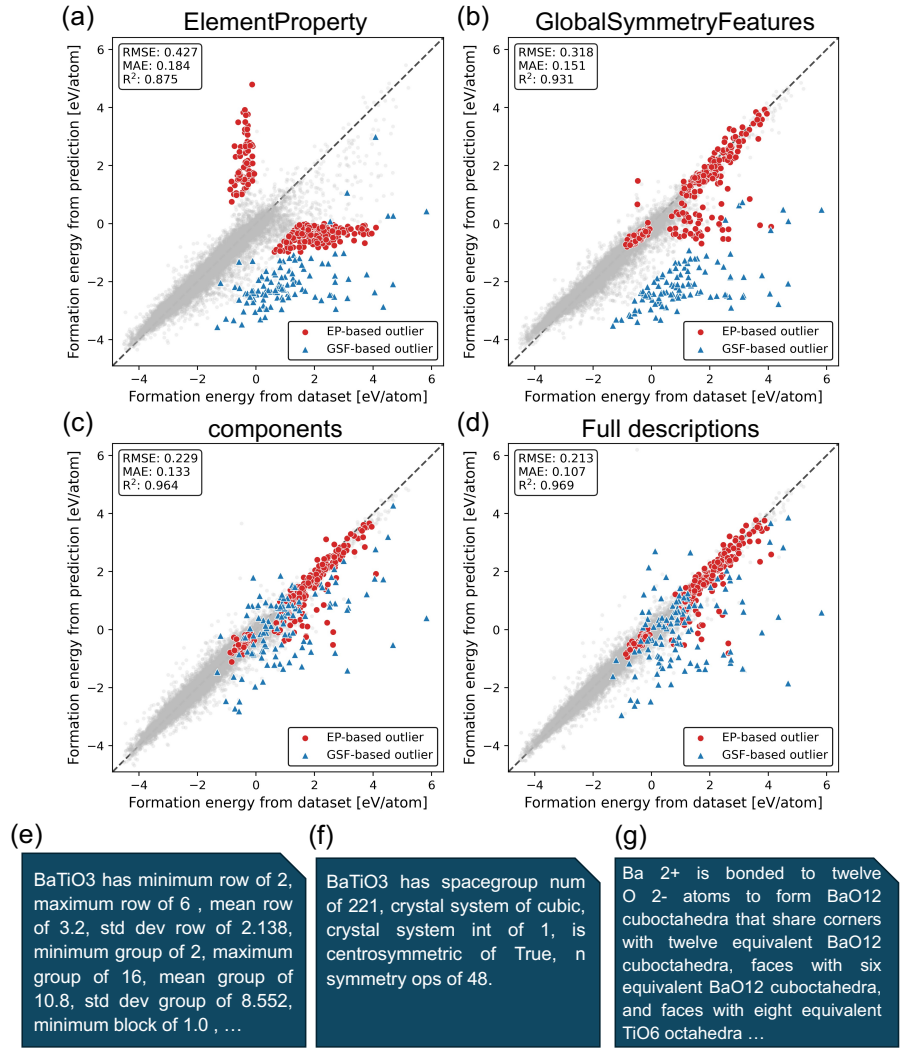

**Figure 5.** Formation energy prediction of the test set on the Textedge dataset. (a)–(d) show parity plots for models trained using (a) ElementProperty only, (b) GlobalSymmetryFeatures only, (c) *components* only, and (d) all available descriptions. (e)–(g) present representative examples of the descriptions for ElementProperty, GlobalSymmetryFeatures, and *components*, respectively.

## 2.3. Utilization of Domain-Specific LLMs

Domain specialization through continual pretraining generally enhances performance on downstream tasks. However, in the original LLM-Prop study, the general-purpose T5 encoder was used as the backbone, and it outperformed the domain-specific MatBERT, which was pretrained on materials-science literature. To further investigate the effect of domain specialization, we compared the general-purpose BERT and the domain-specific MatTPUSciBERT for both ZEBRA-Prop and LLM-Prop.

We first analyze the effect of domain specialization in ZEBRA-Prop, where the backbone language model is used without fine-tuning. As shown in **Table 3** and **Table 4**, all domain-specific LLMs considered in this study, including SciBERT, MatBERT, MatSciBERT, and MatTPUSciBERT, consistently outperform the general-purpose BERT across all target properties on both the in-house and TextEdge datasets. This improvement is observed for every evaluated physical property, indicating that continual pretraining on domain-relevant corpora consistently improves the quality of text embeddings for property prediction.

**Table 3.** Comparison of predictive accuracy between general (BERT) and domain-specific (SciBERT, MatBERT, MatSciBERT, MatTPUSciBERT) LLMs as a backbone model for ZEBRA-Prop on the in-house dataset.

| Backbone LLM | Band gap [eV] | | Formation Energy [eV / atom] | | Dielectric const. (electronic) | | Dielectric const. (ionic) | |
|---|---|---|---|---|---|---|---|---|
| | MAE↓ | $R^2$↑ | MAE↓ | $R^2$↑ | MAE↓ | $R^2$↑ | MAE↓ | $R^2$↑ |
| BERT | 0.615±0.040 | 0.882±0.013 | 0.126±0.007 | 0.954±0.004 | 0.040±0.002 | 0.892±0.010 | 0.128±0.005 | 0.678±0.016 |
| SciBERT | 0.589±0.024 | 0.894±0.006 | 0.111±0.003 | 0.964±0.002 | 0.038±0.002 | 0.901±0.015 | 0.126±0.005 | 0.684±0.007 |
| MatBERT | **0.562±0.039** | **0.897±0.012** | 0.106±0.002 | 0.966±0.003 | **0.036±0.002** | **0.909±0.014** | **0.119±0.004** | **0.716±0.025** |
| MatSciBERT | 0.574±0.039 | 0.895±0.014 | 0.106±0.005 | 0.968±0.003 | 0.037±0.002 | 0.906±0.010 | 0.125±0.003 | 0.694±0.020 |
| MatTPUSciBERT | 0.571±0.032 | 0.894±0.012 | **0.102±0.003** | **0.969±0.002** | 0.038±0.002 | 0.895±0.016 | 0.123±0.007 | 0.693±0.026 |

**Table 4.** Comparison of predictive accuracy between general (BERT) and domain-specific (SciBERT, MatBERT, MatSciBERT, and MatTPUSciBERT) LLMs as a backbone model for ZEBRA-Prop on the TextEdge dataset.

| Backbone LLM | Band gap [eV] | | Formation Energy [eV / atom] | |
|---|---|---|---|---|
| | MAE↓ | $R^2$↑ | MAE↓ | $R^2$↑ |
| BERT | 0.318±0.003 | 0.819±0.002 | 0.111±0.001 | 0.966±0.003 |
| SciBERT | 0.313±0.004 | 0.822±0.004 | 0.107±0.001 | 0.968±0.002 |
| MatBERT | **0.309±0.004** | **0.823±0.007** | **0.106±0.001** | **0.970±0.002** |
| MatSciBERT | 0.310±0.003 | 0.822±0.007 | 0.107±0.001 | 0.968±0.002 |
| MatTPUSciBERT | 0.314±0.005 | 0.820±0.004 | 0.107±0.001 | 0.968±0.002 |

We then examine the impact of domain specialization in LLM-Prop using the in-house dataset, which was not evaluated in the original LLM-Prop study. In addition to the previously reported comparison between the T5 encoder and MatBERT, we include BERT and MatTPUSciBERT under the same evaluation setup. The results in **Table 5** show that both MatBERT and MatTPUSciBERT outperform the general-purpose BERT and also achieve higher predictive accuracy than the T5 encoder on the in-house dataset.

Taken together, these results indicate that domain specialization can improve predictive performance regardless of whether fine-tuning is applied. At the same time, the effectiveness of domain-specific pretraining may depend on dataset characteristics and hyperparameter configurations, suggesting that its benefits could be diminished or even reversed under different evaluation conditions.

**Table 5.** Comparison of predictive accuracy of LLM-Prop using different backbone models (BERT, MatBERT, and MatTPUSciBERT), and a T5 encoder on the in-house dataset. All values are averaged over the test folds in 5-fold cross-validation.

| Backbone LLM | Band gap [eV] | | Formation Energy [eV / atom] | | Dielectric const. (electronic) | | Dielectric const. (ionic) | |
|---|---|---|---|---|---|---|---|---|
| | MAE↓ | $R^2$↑ | MAE↓ | $R^2$↑ | MAE↓ | $R^2$↑ | MAE↓ | $R^2$↑ |
| BERT | 0.506±0.030 | 0.916±0.011 | 0.073±0.003 | 0.982±0.001 | 0.032±0.001 | 0.921±0.013 | 0.111±0.006 | 0.733±0.036 |
| MatBERT | 0.488±0.033 | 0.923±0.011 | 0.077±0.003 | 0.982±0.001 | 0.030±0.001 | 0.931±0.013 | 0.108±0.005 | 0.742±0.025 |
| MatTPUSciBERT | **0.471±0.020** | **0.925±0.005** | **0.069±0.005** | **0.984±0.002** | **0.030±0.002** | **0.932±0.014** | **0.108±0.006** | **0.750±0.028** |
| T5 encoder | 0.522±0.028 | 0.911±0.011 | 0.085±0.005 | 0.978±0.002 | 0.033±0.002 | 0.917±0.018 | 0.117±0.007 | 0.718±0.024 |

## 2.4. Effect of Text Preprocessing

In the original LLM-Prop study, text preprocessing consisted of replacing numerical values appearing in the text with dedicated tokens, and removing stopwords that were considered not to directly contribute to prediction performance. The replacement of numerical values was introduced to address the difficulty of LLMs in handling numerical information. However, in materials property prediction tasks, numerical values often carry essential physical meaning through their magnitudes and relative relationships. From this perspective, simply discarding numerical information, as in LLM-Prop, may be suboptimal. Accordingly, we designed alternative preprocessing strategies for chemical formula representations and numerical values. Specifically, chemical formulas are converted into simplified representations that remove parentheses and describe only element symbols and their stoichiometric ratios, making them more interpretable for LLMs. Numerical values are transformed through scaling and subsequent rounding to integers, while preserving the original numerical information. Unlike approaches such as LLM-Prop, which rely on replacing numerical values, our preprocessing strategies aim to improve model tractability without reducing the semantic information contained in materials-related text. Details of each preprocessing method are provided in Section 4.3.

**Table 6** and **Table 7** present the results of an ablation study evaluating the effects of these preprocessing strategies on model performance. To ensure a precise comparison of text preprocessing strategies, this evaluation was conducted using only matminer descriptions. The results show that both the conversion to simplified chemical formulas and the integerization of numerical values individually lead to performance improvements. Moreover, their combined use yields a synergistic effect beyond that achieved by either strategy alone. These findings support the approach of leveraging numerical information through appropriate rescaling rather than discarding it, and provide practical guidance for designing textual representations in materials science.

**Table 6.** Dependence of ZEBRA-Prop prediction accuracy on text preprocessing strategies on the in-house dataset, with values averaged over the test folds in 5-fold cross-validation.

| Text preprocessing | Band gap [eV] | | Formation Energy [eV / atom] | | Dielectric const. (electronic) | | Dielectric const. (ionic) | |
|---|---|---|---|---|---|---|---|---|
| | MAE↓ | $R^2$↑ | MAE↓ | $R^2$↑ | MAE↓ | $R^2$↑ | MAE↓ | $R^2$↑ |
| ZEBRA-Prop (w/o robocrys.) | **0.599±0.025** | **0.882±0.011** | **0.103±0.003** | **0.966±0.003** | **0.042±0.003** | 0.880±0.013 | **0.138±0.007** | **0.635±0.025** |
| - simplified chemical formula | 0.638±0.052 | 0.869±0.022 | 0.115±0.003 | 0.959±0.005 | 0.042±0.003 | **0.881±0.016** | 0.141±0.006 | 0.616±0.032 |
| - integerization of numerical values | 0.645±0.032 | 0.867±0.014 | 0.120±0.004 | 0.957±0.005 | 0.046±0.001 | 0.865±0.012 | 0.142±0.006 | 0.621±0.016 |
| - both | 0.690±0.053 | 0.852±0.018 | 0.132±0.005 | 0.951±0.005 | 0.048±0.003 | 0.853±0.013 | 0.146±0.009 | 0.599±0.040 |

**Table 7.** Dependence of ZEBRA-Prop prediction accuracy on text preprocessing strategies on the TextEdge dataset, with values averaged over the test folds in 5-fold cross-validation.

| Text preprocessing | Band gap [eV] | | Formation Energy [eV / atom] | |
|---|---|---|---|---|
| | MAE↓ | $R^2$↑ | MAE↓ | $R^2$↑ |
| ZEBRA-Prop (w/o robocrys.) | **0.345±0.004** | **0.784±0.004** | **0.123±0.002** | **0.939±0.007** |
| - simplified chemical formula | 0.346±0.006 | 0.784±0.005 | 0.124±0.003 | 0.939±0.007 |
| - integerization of numerical values | 0.360±0.007 | 0.765±0.007 | 0.131±0.002 | 0.939±0.007 |
| - both | 0.371±0.005 | 0.754±0.009 | 0.137±0.002 | 0.929±0.007 |

## 3. Conclusion

In this study, we proposed an efficient LLM-based material property prediction model that does not require fine-tuning and demonstrated its effectiveness through empirical validation. It achieved a promising balance between computational efficiency and predictive accuracy, showing approximately a 95% reduction in training time compared with LLM-Prop while maintaining comparable predictive performance. By leveraging MatTPUSciBERT, the model effectively incorporated domain-specific knowledge tailored to materials science. In addition, the integration of diverse textual descriptions through weighted aggregation allowed the model to overcome the context-length limitations inherent to LLMs and further improved predictive accuracy.

The predictive performance of ZEBRA-Prop is expected to further improve with ongoing advances in foundation LLMs and domain-specific LLMs, as well as through the enhancement of input text quality and diversity. Although GNN-based models such as CGCNN and ALIGNN showed superior predictive accuracy in this study, their application generally relies on the availability of well-defined crystal structures that can be explicitly represented as graphs. In contrast, experimental data often include free-text or semi-structured information, such as synthesis conditions, heat-treatment histories, measurement settings, and observational notes, and may also involve the effects of lattice defects, mixed phases, or non-equilibrium states that are difficult to represent as a single well-defined crystal graph. From this perspective, directly applying GNN-based frameworks to experimental data is not always straightforward. By contrast, the text-based nature and architectural generality of ZEBRA-Prop provide a potentially flexible route for incorporating such experimentally derived information. In principle, the proposed framework can extend beyond computational crystal-structure descriptions to directly utilize laboratory notebooks, synthesis records, and other experimental documents as model inputs[35]. Such an extension may enable prediction of experimentally measured properties, thereby broadening the applicability of the framework from computational materials screening to practical support of experiments. Moreover, given the architectural generality of the proposed framework, extensions to CLIP[36]-like multimodal models represent an attractive direction for future research. Through these developments, the proposed method is expected to evolve into a practical tool for knowledge discovery and materials design in materials science.

## 4. Methods
### 4.1. Architecture of ZEBRA-Prop

**Figure 6** illustrates the architectures of ZEBRA-Prop and LLM-Prop. ZEBRA-Prop builds upon LLM-Prop while addressing its key limitations, differing in the following three aspects.

First, ZEBRA-Prop keeps the parameters of the LLM frozen during training, whereas LLM-Prop fine-tunes the entire LLM. This design substantially reduces the number of trainable parameters, leading to a lower training cost.

Second, ZEBRA-Prop processes multiple short texts, each fitting within the LLM's context window, whereas LLM-Prop relies on a single long text. The original LLM-Prop paper reports that approximately 20% of the input texts were truncated due to context-length limitations, which may have resulted in information loss and degraded predictive performance. The embeddings obtained from these texts are then integrated through a weighted summation mechanism. This strategy mitigates the negative impact of context-length constraints observed in LLM-Prop. Such an approach is difficult to adopt in fine-tuning–based models such as LLM-Prop, because handling multiple independent inputs would substantially increase the computational cost.

Third, ZEBRA-Prop adopts a different text preprocessing strategy from that of LLM-Prop. While LLM-Prop removes stopwords to compress the input context and replaces numerical values with [NUM] and [ANG] tokens, learning their semantics through fine-tuning, ZEBRA-Prop applies preprocessing designed to make numerical values and chemical symbols easier for the LLM to handle directly. Specifically, chemical formulas are converted into simplified chemical formulas, and numerical values are transformed into integerized representations.

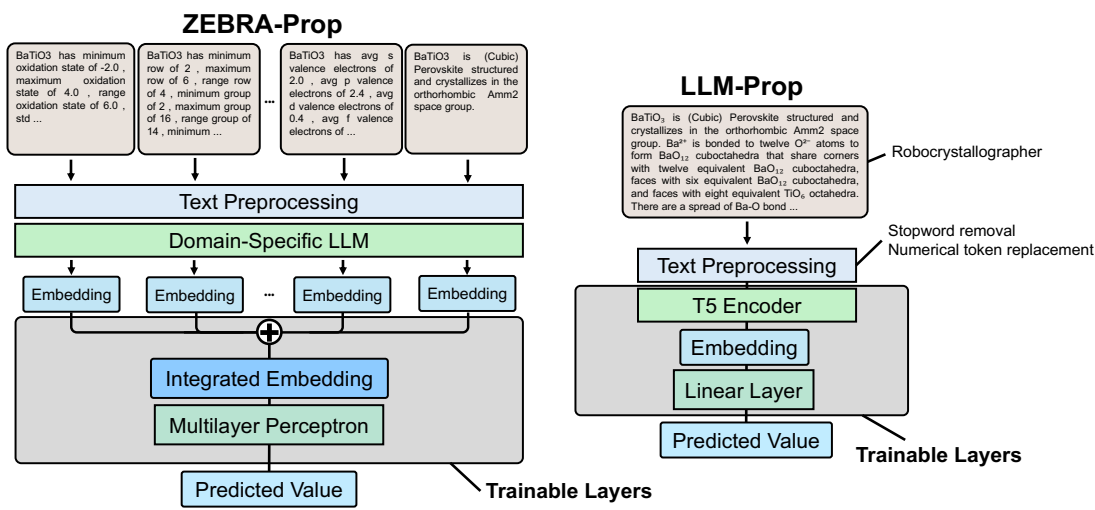

**Figure 6.** Architectural comparison between ZEBRA-Prop and LLM-Prop.

### 4.2. Input Text Generation

In general, material properties are determined by a complex interplay of various factors related to a material's chemical composition and atomic arrangement. In the LLM-Prop framework, textual inputs are generated exclusively using Robocrystallographer, a tool that automatically produces natural-language descriptions of crystal structures based on a single predefined format. The descriptions generated by Robocrystallographer consist of three parts: *mineral*, which summarizes the overall crystal structure using the mineral prototype and space group; *component makeup*, which describes the dimensionality, number, and orientation of structural components; and *components*, which provides detailed information on the atomic-site coordination, bonding, and polyhedral connectivity.

To evaluate whether incorporating multiple perspectives improves predictive performance and to analyze the effect of the weight integration mechanism in ZEBRA-Prop, we adopt a framework capable of incorporating arbitrary user-defined textual descriptions. Specifically, in addition to the crystal-structure descriptions generated by Robocrystallographer, we automatically generate complementary textual descriptions using descriptors obtained from the matminer code, a materials data-mining library. These descriptors capture diverse aspects of materials, including chemical composition and crystal structure, thereby enabling the integration of multiple descriptive perspectives.

The descriptions used for evaluation in this study are constructed from descriptors generated by ten featurizers in matminer mapped into a predefined textual format: "*[chemical formula] has [name of descriptor 1] of [value of descriptor 1], [name of descriptor 2] of [value of descriptor 2]…*". When a descriptor contains missing values, the corresponding positions are replaced with a [MASK] token. The fraction of data samples with missing values in the matminer descriptions for each dataset is shown in **Table 8**. The matminer featurizers used to generate the descriptors are listed in **Table S1**, and representative examples of the generated descriptors are provided in **Table S2** (Supporting Information). In addition to these ten descriptions, we also evaluate a setting in which the *mineral* and *components* descriptions generated by Robocrystallographer are included. Although Robocrystallographer can also generate *component makeup* descriptions, these are available only for a limited subset of the data. For this reason, we do not include *component makeup* descriptions in the evaluation.

**Table 8.** Proportion of data containing missing values for each matminer featurizer in the in-house and TextEdge datasets.

| Featurizer | In-house (2,202 data) | | TextEdge (138,378 data) | |
| --- | --- | --- | --- | --- |
| | Count | Ratio (%) | Count | Ratio (%) |
| AtomicOrbitals | 0 | 0.00 | 1,666 | 1.20 |
| AtomicPackingEfficiency | 0 | 0.00 | 1,306 | 0.94 |
| DensityFeatures | 0 | 0.00 | 167 | 0.12 |
| ElectronegativityDiff | 0 | 0.00 | 27,311 | 19.74 |
| ElementFraction | 0 | 0.00 | 0 | 0.00 |
| ElementProperty | 0 | 0.00 | 167 | 0.12 |
| GlobalSymmetryFeatures | 0 | 0.00 | 0 | 0.00 |
| IonProperty | 0 | 0.00 | 2 | < 0.01 |
| OxidationStates | 0 | 0.00 | 0 | 0.00 |
| ValenceOrbital | 0 | 0.00 | 0 | 0.00 |

**4.3. Text Preprocessing**

*4.3.1 Conversion to Simplified Chemical Formulas*

Standard chemical notation uses parentheses to denote grouped polyatomic units and their multiplicities (e.g., $Ca_{10}(PO_4)_6(OH)_2$). While such notation is intuitive for scientists, it is not necessarily optimal for LLMs. To address this issue, several prior studies have explored approaches based on chemical formula standardization. For example, Park *et al.* adopted a *reduced composition* format to provide stoichiometric information for text−structure alignment and text-guided generation[10].

In this study, we preprocess chemical formulas by explicitly expanding all elements and their corresponding quantities and replacing the original representation with a simplified chemical formula in which elements are listed sequentially. Unlike *reduced composition*, this representation does not apply division by the greatest common divisor of element counts. For example, $Cu(NO_3)_2$ is converted to Cu 1 N 2 O 6 by expanding the grouped units and enumerating each element with its explicit count.

*4.3.2 Integerization of Numerical Values*

It has been widely reported that LLMs have difficulty processing numerical information effectively[37,38], primarily because they treat numbers as discrete textual tokens rather than continuous quantities. In particular, floating-point values and values with extremely large or small magnitudes are difficult for LLMs to handle, as they appear infrequently in training corpora. In the context of property prediction tasks within the materials science domain, understanding numerical magnitudes of descriptors and their relative relationships is particularly critical, especially for models without fine-tuning, which must rely on the numerical representations acquired during pretraining.

To address this issue, our model transforms numerical values in the input text into integers through a scaling strategy that adjusts the overall numerical scale of the dataset. This preprocessing step is designed to enhance numerical reasoning by aligning numerical inputs with the operational characteristics of language models. While LLM-Prop replaces numerical values with special tokens, our approach preserves the original numerical information through scaling, making values easier for LLMs to process.

In the evaluations presented in this study, the numerical scale is determined systematically according to a predefined procedure. The specific scaling procedure is as follows. First, we compute the mean of the absolute values across the dataset. Next, all values are multiplied by $10^n$, where $n$ is an integer chosen such that the resulting mean falls within the range of 10 to 100. For example, if the mean absolute value of a descriptor is 0.3, all values are multiplied by 100, whereas if the mean absolute value is 300,000, all values are multiplied by 0.0001.

In practical applications, the scaling factor can also be chosen based on domain knowledge. Considering the typical order of magnitude and physical meaning of the target quantity may lead to a more appropriate preprocessing strategy.

### 4.4. Weighted Summation Mechanism

The embeddings obtained from multiple textual descriptions are integrated using a weighted summation mechanism, defined as follows:

Let $D$ denote the number of input textual descriptions and $H$ the dimensionality of each description embedding. We represent the embeddings corresponding to the descriptions as $\{e_d\}_{d=1}^{D}$, where $e_d \in \mathbb{R}^H$.

The integrated representation $z \in \mathbb{R}^H$ is computed as

$$z = \sum_{d=1}^{D} w_d e_d \tag{1}$$

where $w = (w_1, \cdots, w_D)$ is a learnable weight vector shared across samples.

This formulation enables the aggregation of information from multiple independent textual inputs while preserving a fixed embedding dimensionality $H$. Notably, even as the number of input descriptions $D$ increases, the training and inference costs of the prediction head remain unchanged. In our implementation, the weights $w$ are not constrained to the probability simplex (e.g., via softmax normalization), nor are nonlinear activation functions applied.

### 4.5. Evaluation Setup

*4.5.1 Datasets*

In this study, we employed two distinct datasets to evaluate the proposed method.

The first dataset, referred to as a subset of the TextEdge dataset, is derived from the original TextEdge dataset, which has previously been used to evaluate the LLM-Prop model. The original TextEdge dataset contains 144,931 entries and includes target properties obtained from the Materials Project database, along with natural language descriptions generated by Robocrystallographer for each material. In this work, we introduced the formation energy as an additional target property, which was not included in the original TextEdge dataset. During this process, we re-queried the Materials Project database (accessed in January 2026), which had been updated since the release of the original dataset. As a result of the database updates and the removal of entries with missing values in any of the target properties, the dataset size was reduced from 144,931 to 138,378 entries.

In addition to the TextEdge dataset, we also utilized an original in-house dataset constructed by our research group based on first-principles calculations. This dataset comprises non-magnetic materials selected from the Materials Project database that contain at least one of the following elements: oxygen (O), sulfur (S), or selenium (Se). For each target material, properties were computed using first-principles methods, including the band gap, formation energy, and both the electronic and ionic contributions to the dielectric constant. The detailed computational procedures used to construct this in-house dataset are described in Section 4.5.2.

The distribution of the target properties is shown in **Figure S1** (Supporting Information).

*4.5.2 Computational Details*

The in-house dataset is drawn from the same database as our previous study[39]. The computational workflow for constructing this dataset is illustrated in **Figure 7**.

First, we retrieved 14,122 oxides and chalcogenides from the Materials Project database. Materials were pre-screened according to the criteria described in Ref.[39]: briefly, non-magnetic oxides and chalcogenides (containing O, S, or Se) excluding certain elements (H, halogens, noble gases, etc.) and complex structures (space group *P*1 or >40 atoms per primitive cell). All first-principles calculations were performed using the projector augmented-wave method[40] as implemented in the VASP code[41,42] (version 5.4.4). Band structures were computed using both the Perdew–Burke–Ernzerhof functional tuned for solids (PBEsol)[43] (+$U$[44,45]) and the non-self-consistent dielectric-dependent (nsc-dd) hybrid functional approach[46,47], and dielectric constants (ionic and electronic contributions) were obtained from density functional perturbation theory (DFPT)[48,49] calculations. Materials were included in the final dataset if they satisfied the following criteria: (1) thermodynamically stable, defined by an energy above the convex hull of 0 eV, (2) a nsc-dd hybrid band gap of at least 0.5 eV, (3) no optical phonon frequencies lower than 0.1 THz, and (4) no acoustic phonon modes with imaginary frequencies at the Γ point whose absolute values exceed 0.3 THz. We employed effective *U* parameters of 3 eV for the *d* states of Ti, V, Cr, Zr, Nb, Mo, Pd, Hf, Ta, W, Re, Pt, and Au, and 5 eV for the *d* states of Cu, Zn, Ag, and the *f* states of Ce.

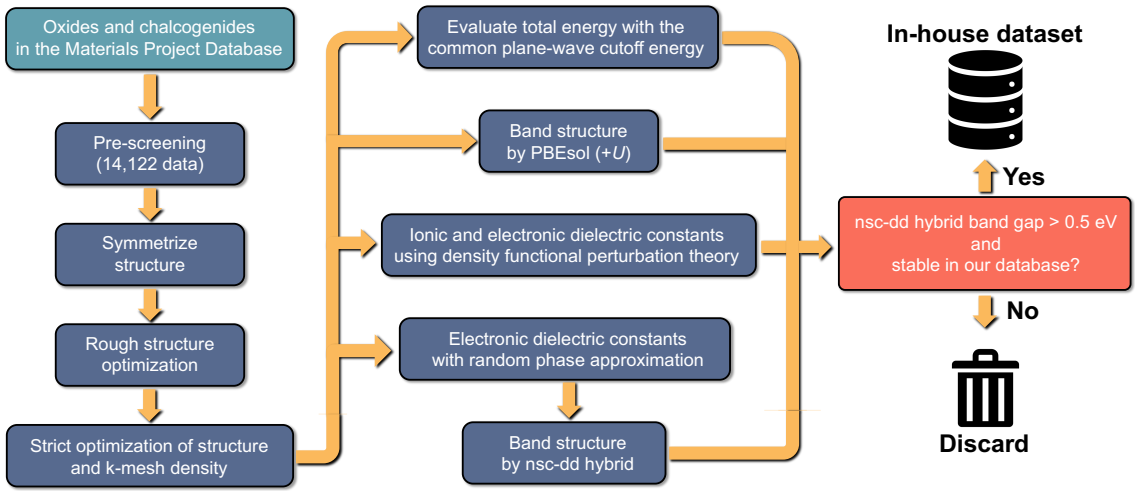

**Figure 7.** The computational workflow for constructing in-house dataset.

*4.5.3 Evaluation Environment*

In our evaluations, we utilized node_q, a logical partition representing one-fourth of a single TSUBAME4.0 compute node at Institute of Science Tokyo. Each TSUBAME4.0 node is equipped with two AMD EPYC 9654 processors (2.4 GHz, 96 cores / 192 threads each), totaling 192 cores / 384 threads, 768 GiB of DDR5-4800 memory, and four NVIDIA H100 SXM5 GPUs with 94 GiB of HBM2e memory per GPU. Accordingly, a single node_q provides access to one GPU, 48 CPU cores (96 threads), and 192 GiB of memory.

*4.5.4 Evaluation Setup*

As evaluation metrics to assess model performance in property prediction tasks, we used the mean absolute error (MAE) and the coefficient of determination ($R^2$). To ensure robustness and generalizability, the dataset was partitioned using 5-fold cross-validation with random shuffling. The data were divided into five approximately equal-sized folds, with any remainder distributed evenly so that fold sizes differed by at most one sample. For each cross-validation iteration, one fold was designated as the test set, the subsequent fold (in cyclic order) was used as the validation set, and the remaining three folds were combined to form the training set. This procedure ensured that the training, validation, and test sets were mutually exclusive within each iteration. The same cross-validation splits were used for all models to guarantee fair comparison.

The ZEBRA-Prop model was trained using the following hyperparameters. The learning rate was set to 0.001 and the batch size was 64. The MLP consisted of three hidden layers with sizes [128, 256, 512], and a dropout rate of 0.2 was applied to each layer. The learning rate and batch size were determined based on a grid search, the results of which are provided in **Figures S8–S12** (Supporting Information).

For comparison, we evaluated ZEBRA-Prop against several baseline models, including LLM-Prop, CGCNN, ALIGNN, and a random forest model trained using the same set of descriptors derived from the matminer code as those used to construct the input features for ZEBRA-Prop.

For CGCNN, we employed the default architecture and training configuration provided in the original implementation, with a learning rate of 0.01, a batch size of 32, 200 training epochs, and a graph convolutional depth of 3. Similarly, ALIGNN was trained using its default settings, with a learning rate of 0.001, a batch size of 32, and 300 training epochs.

For LLM-Prop, we followed the training protocol of the original implementation. The encoder-only variant of the T5-small model was fine-tuned using a learning rate of $1\times10^{-3}$, a dropout rate of 0.2, and the Adam optimizer with a one-cycle learning rate scheduler. The model was trained for 200 epochs using 888 input tokens with a batch size of 64, and model selection was based on validation performance. In addition, we trained the BERT-based model (BERT, MatBERT, and MatTPUSciBERT) using their respective tokenizers with a learning rate of $5\times10^{-5}$, a batch size of 64, and a dropout rate of 0.5 for 200 epochs, employing the Adam optimizer with a one-cycle learning rate scheduler, following the original implementation.

ZEBRA-Prop, LLM-Prop, CGCNN, and ALIGNN were trained on the training sets of the 5-fold cross-validation setup. For each model, the epoch that achieved the lowest MAE on the corresponding validation set was selected, and the resulting model was used to evaluate predictive performance on the test set.

The random forest model was trained using the implementation provided by the scikit-learn[50] library. A manual grid search over n_estimators = [200, 400, 800] and max_depth = [None, 10, 20] was performed by fitting each parameter combination once on the training split of each fold and selecting the model with the lowest validation MAE. The selected model was not retrained; it was evaluated directly on the training, validation, and test sets, and the test performance was reported.

*4.5.5 Implementation Details*

We implemented ZEBRA-Prop in Python (version 3.12.12) using the following libraries: transformers[51] (version 5.0.0), PyTorch[52] (version 2.10.0), matminer[31] (version 0.10.0), and pymatgen[53] (version 2025.10.7).

The transformers and PyTorch codes were used for model construction and training, while the matminer and pymatgen codes were employed for materials data processing and feature extraction. All evaluations were conducted using the same software environment to ensure reproducibility.

**4.6 Use of Artificial Intelligence Generated Content**

ChatGPT (OpenAI) and Claude (Anthropic) were used for language refinement to improve clarity and grammar. OpenAI Codex was used to assist in code refactoring. All outputs were reviewed and approved by the authors.

**5. Conflict of Interest**

The authors declare no conflict of interest.

**6. Data Availability Statement**

The implementation of ZEBRA-Prop is available at: https://github.com/oba-group/ZEBRA-Prop. The in-house dataset used for evaluation is publicly available at: https://github.com/oba-group/ZEBRA-dataset.

**7. Acknowledgments**

This work was supported by KAKENHI (Grant Numbers: JP24K08562 and JP24H00376) from the Japan Society for the Promotion of Science, Data Creation and Utilization Type Material Research and Development Project (Grant Number: JPMXP1122683430) and Design & Engineering by Joint Inverse Innovation for Materials Architecture Project from the Ministry of Education, Culture, Sports, Science, and Technology, Japan, and the KISTEC Project.

Computations were carried out using the TSUBAME4.0 supercomputer at Institute of Science Tokyo.

Supporting Information: ZEBRA-Prop: A Zero-Shot Embedding-Based Rapid and Accessible Regression Model for Materials Properties

Ryoma Yamamoto[1*], Akira Takahashi[1*], Kei Terayama[2,3], Yu Kumagai[4], Fumiyasu Oba[1,3,5]

1. Materials and Structures Laboratory, Institute of Integrated Research, Institute of Science Tokyo, 4259 Nagatsuta, Midori-ku, Yokohama 226-8501, Japan
2. Graduate School of Medical Life Science, Yokohama City University, 1-7-29 Suehiro-cho, Tsurumi-ku, Yokohama 230-0045, Japan
3. MDX Research Center for Element Strategy, Institute of Integrated Research, Institute of Science Tokyo, 4259 Nagatsuta, Midori-ku, Yokohama 226-8501, Japan
4. Institute for Materials Research, Tohoku University, 2-2-1 Katahira, Aoba-ku, Sendai 980-8577, Japan
5. Kanagawa Institute of Industrial Science and Technology (KISTEC), 705-1 Shimoimaizumi, Ebina 243-0435, Japan

*E-mail: yamamoto.r.ff57@m.isct.ac.jp (R. Yamamoto), takahashi.a.f9db@m.isct.ac.jp (A. Takahashi)

## 1. Distribution of the Target Properties

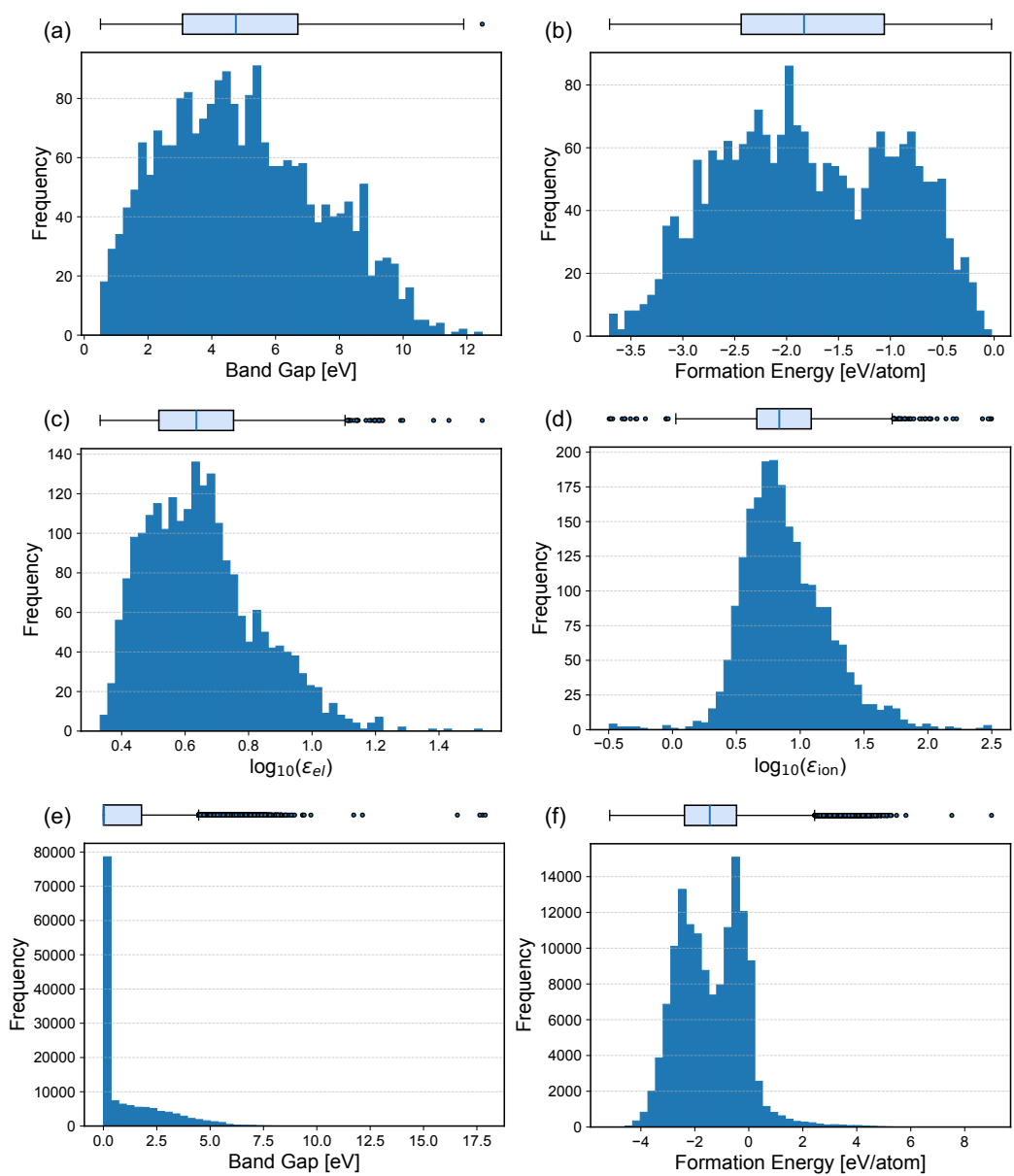

**Figure S1.** Distribution of the target properties: (a) the band gap in the in-house dataset, (b) the formation energy in the in-house dataset, (c) the electronic contribution to the dielectric constant ($\log_{10}$ scale) in the in-house dataset, (d) the ionic contribution to the dielectric constant ($\log_{10}$ scale) in the in-house dataset, (e) the band gap in the TextEdge dataset, and (f) the formation energy in the TextEdge dataset.

## 2. Features of Compounds

**Table S1.** Classes of matminer used to generate features for constructing the input crystal structure descriptions in this study.

| Submodule | Class | Note |
|---|---|---|
| composition | AtomicOrbitals[1] | |
| composition | AtomicPackingEfficiency[2] | |
| composition | ElementFraction | |
| composition | ElementProperty[3] | data_source = ("pymatgen"); features = ["row", "group", "block", "atomic_mass", "atomic_radius"]; stats = ["minimum", "maximum", "mean", "std_dev"] |
| composition | IonProperty[4] | |
| composition | ValenceOrbital[4,5] | |
| composition | ElectronegativityDiff[5] | |
| composition | OxidationStates[5] | |
| structure | DensityFeatures | |
| structure | GlobalSymmetryFeatures | |

## 3. Examples of Crystal Structure Descriptions

**Table S2.** Examples of crystal structure descriptions for BaTiO$_3$. All descriptions are in their original form and have not undergone any text preprocessing. (mp-id: mp-5777, space group: Amm2).

| Description | Example |
| --- | --- |
| AtomicOrbitals | BaTiO3 has HOMO character of p, HOMO element of O, HOMO energy of -0.338381, LUMO character of d, LUMO element of Ti, LUMO energy of -0.17001, gap AO of 0.168371. |
| AtomicPackingEfficiency | BaTiO3 has mean simul. packing efficiency of -0.0055, mean abs simul. packing efficiency of 0.0823, dist from 1 clusters \|APE\| < 0.010 of 0.0537, dist from 3 clusters \|APE\| < 0.010 of 0.0736, dist from 5 clusters \|APE\| < 0.010 of 0.0942. |
| ElementFraction | BaTiO3 has O of 0.6, Ti of 0.2, Ba of 0.2. |
| ElementProperty | BaTiO3 has minimum row of 2, maximum row of 6 , mean row of 3.2, std dev row of 2.138, minimum group of 2, maximum group of 16, mean group of 10.8, std dev group of 8.552, minimum block of 1.0 , maximum block of 3.0, mean block of 2.0, std dev block of 0.845, minimum atomic mass of 16.000 , maximum atomic mass of 137.327, mean atomic mass of 46.638 , std dev atomic mass of 62.798, minimum atomic radius of 0.6 , maximum atomic radius of 2.15, mean atomic radius of 1.070 , std dev atomic radius of 0.832. |
| IonProperty | BaTiO3 has compound possible of True, max ionic char of 0.80321, avg ionic char of 0.17173. |
| ValenceOrbital | BaTiO3 has avg s valence electrons of 2.0, avg p valence electrons of 2.4, avg d valence electrons of 0.4, avg f valence electrons of 0.0, frac s valence electrons of 0.417, frac p valence electrons of 0.5, frac d valence electrons of 0.084, frac f valence electrons of 0.0. |
| ElectronegativityDiff | BaTiO3 has minimum EN difference of 1.9, maximum EN difference of 2.55, range EN difference of 0.65 , mean EN difference of 2.23, std dev EN difference of 0.46. |
| OxidationStates | BaTiO3 has minimum oxidation state of -2.0 , maximum oxidation state of 4.0, range oxidation state of 6.0, std dev oxidation state of 3.38. |
| DensityFeatures | BaTiO3 has density of 5.98, vpa of 12.96, packing fraction of 0.862. |
| GlobalSymmetryFeatures | BaTiO3 has spacegroup num of 221, crystal system of cubic, crystal system int of 1, is centrosymmetric of True, n symmetry ops of 48. |
| Robocrystallographer | BaTiO3 is (Cubic) Perovskite structured and crystallizes in the orthorhombic Amm2 space group. Ba2+ is bonded to twelve O2- atoms to form BaO12 cuboctahedra that share corners with twelve equivalent BaO12 cuboctahedra, faces with six equivalent BaO12 cuboctahedra, and faces with eight equivalent TiO6 octahedra. There are a spread of Ba-O bond distances ranging from 2.79-2.96 Å. Ti4+ is bonded to six O2- atoms to form TiO6 octahedra that share corners with six equivalent TiO6 octahedra and faces with eight equivalent BaO12 cuboctahedra. The corner-sharing octahedral tilt angles range from 6-10°. There are a spread of Ti-O bond distances ranging from 1.87-2.19 Å. There are two inequivalent O2- sites. In the first O2- site, O2- is bonded in a 2-coordinate geometry to four equivalent Ba2+ and two equivalent Ti4+ atoms. In the second O2- site, O2- is bonded in a distorted linear geometry to four equivalent Ba2+ and two equivalent Ti4+ atoms. |
| Mineral (from Robocrystallographer) | BaTiO3 is (Cubic) Perovskite structured and crystallizes in the orthorhombic Amm2 space group. |
| Components (from Robocrystallographer) | Ba2+ is bonded to twelve O2- atoms to form BaO12 cuboctahedra that share corners with twelve equivalent BaO12 cuboctahedra, faces with six equivalent BaO12 cuboctahedra, and faces with eight equivalent TiO6 octahedra. There are a spread of Ba-O bond distances ranging from 2.79-2.96 Å. Ti4+ is bonded to six O2- atoms to form TiO6 octahedra that share corners with six equivalent TiO6 octahedra and faces with eight equivalent BaO12 cuboctahedra. The corner-sharing octahedral tilt angles range from 6-10°. There are a spread of Ti-O bond distances ranging from 1.87-2.19 Å. There are two inequivalent O2- sites. In the first O2- site, O2- is bonded in a 2-coordinate geometry to four equivalent Ba2+ and two equivalent Ti4+ atoms. In the second O2- site, O2- is bonded in a distorted linear geometry to four equivalent Ba2+ and two equivalent Ti4+ atoms. |

## 4. Parity Plots for Each Model

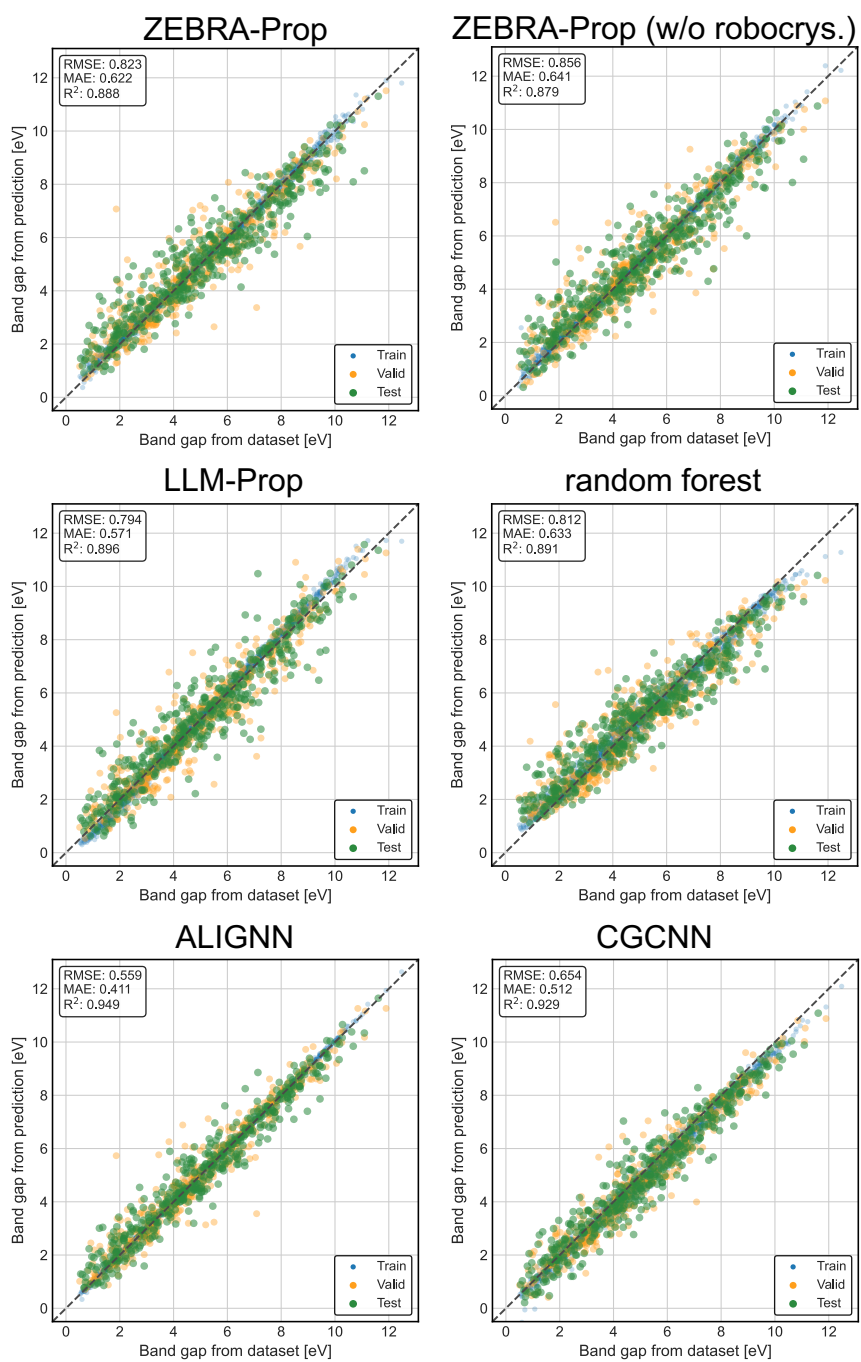

**Figure S2.** Predicted values with respect to values of the in-house dataset for the train, validation, and test sets.

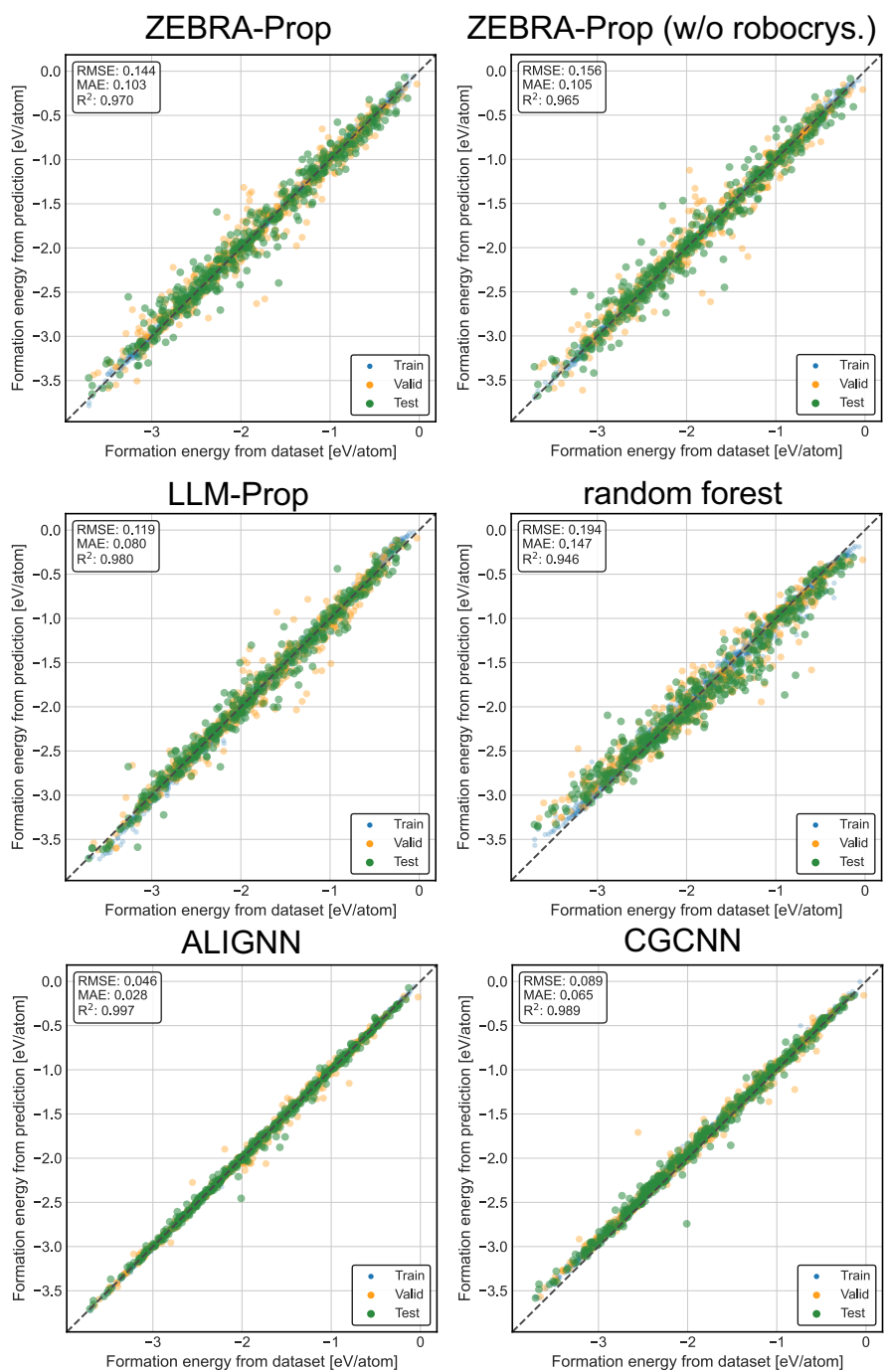

**Figure S2**(Continued).

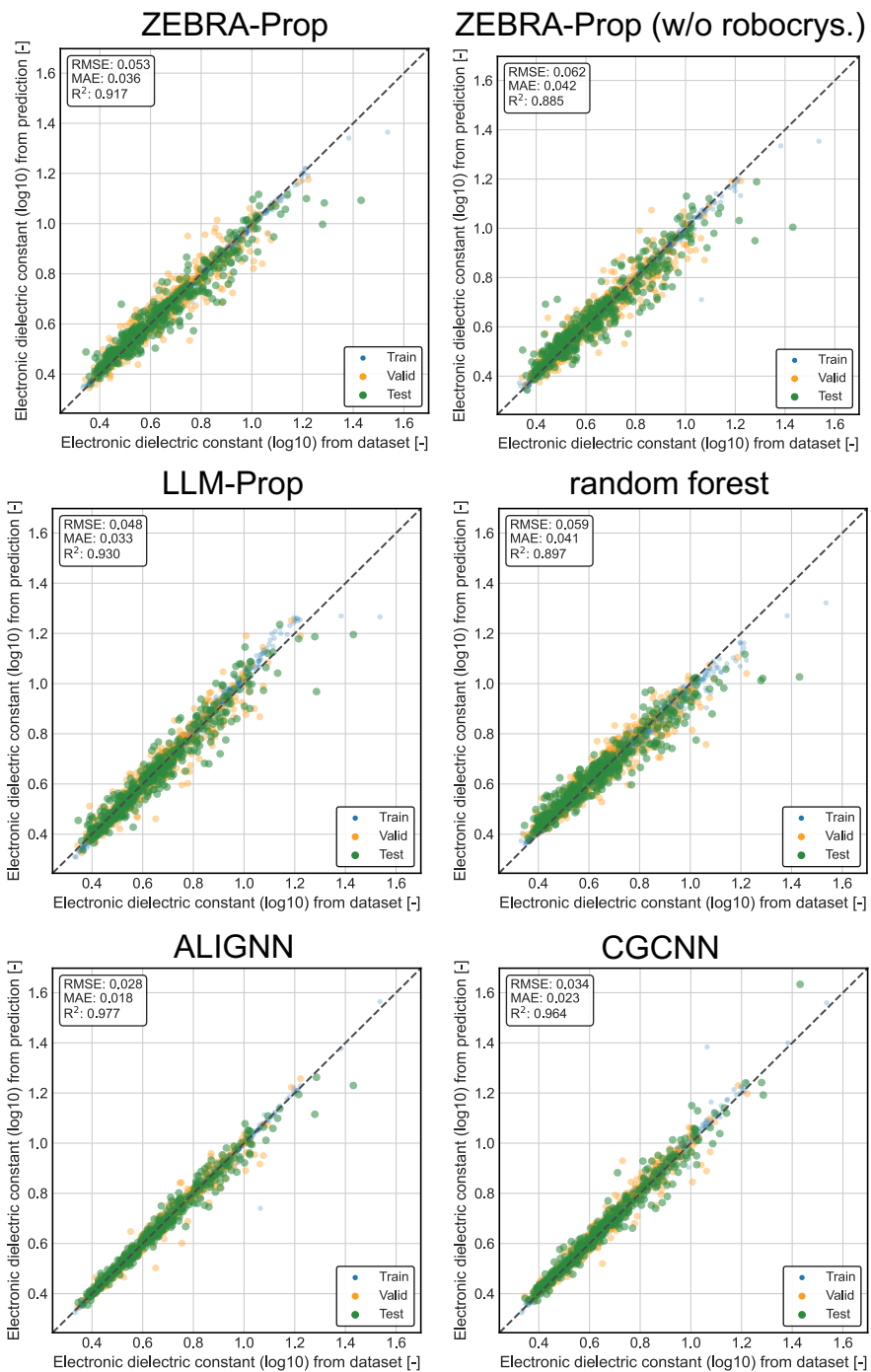

**Figure S2**(Continued).

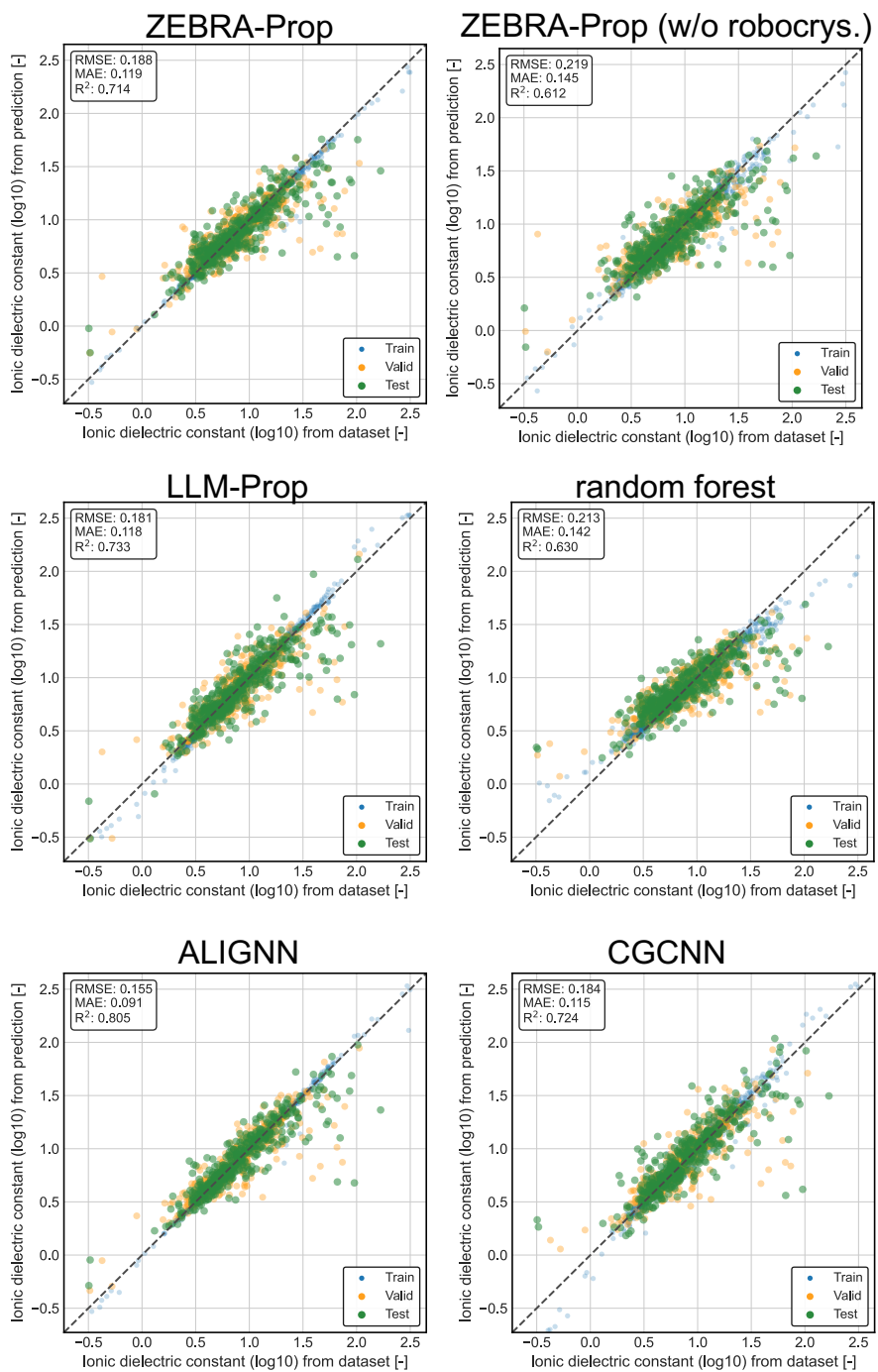

**Figure S2**(Continued)**.**

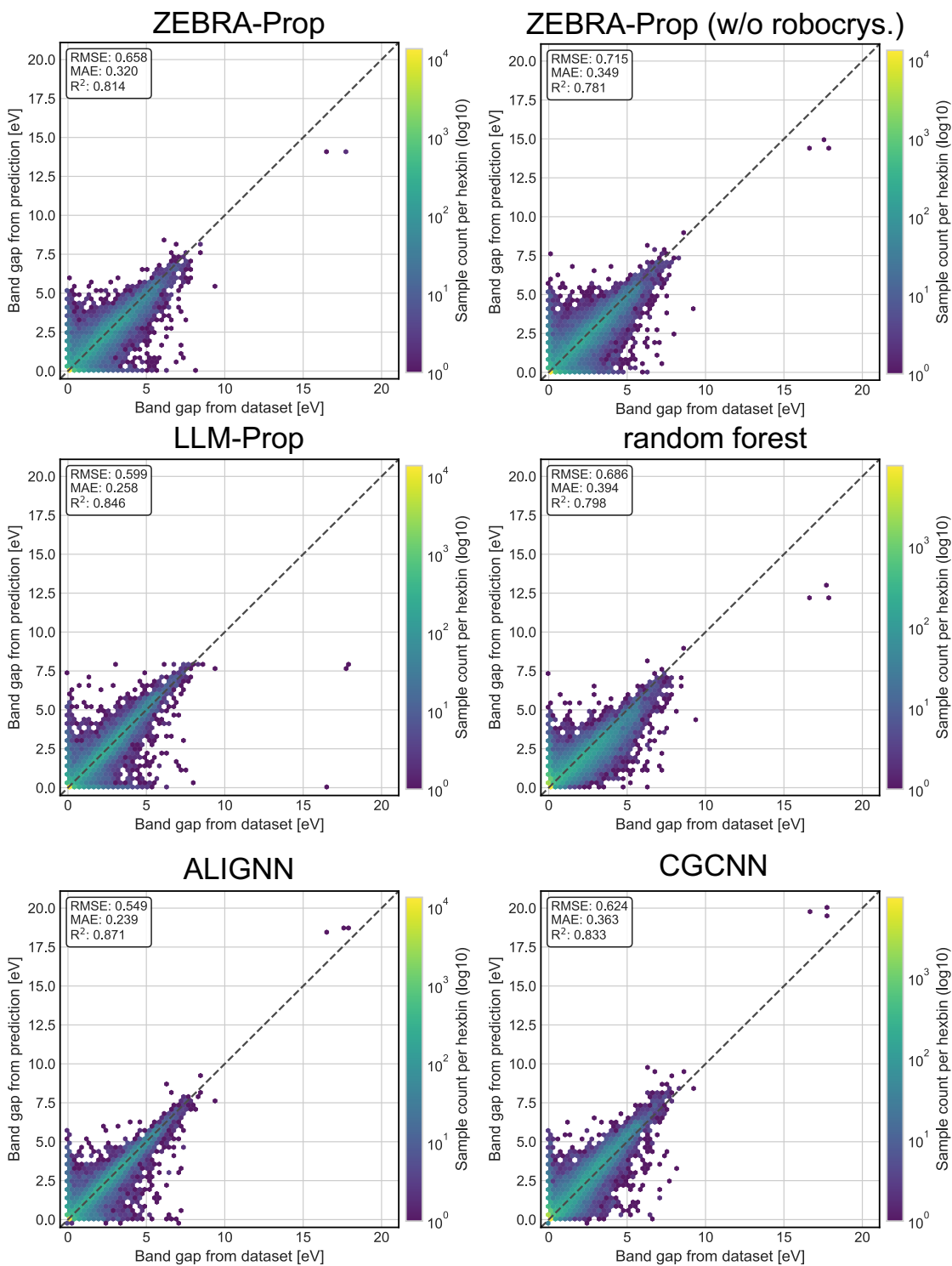

**Figure S3.** Hexbin map plot of the predicted values with respect to values of the TextEdge dataset for the test set, where color indicates sample count per hexagonal bin.

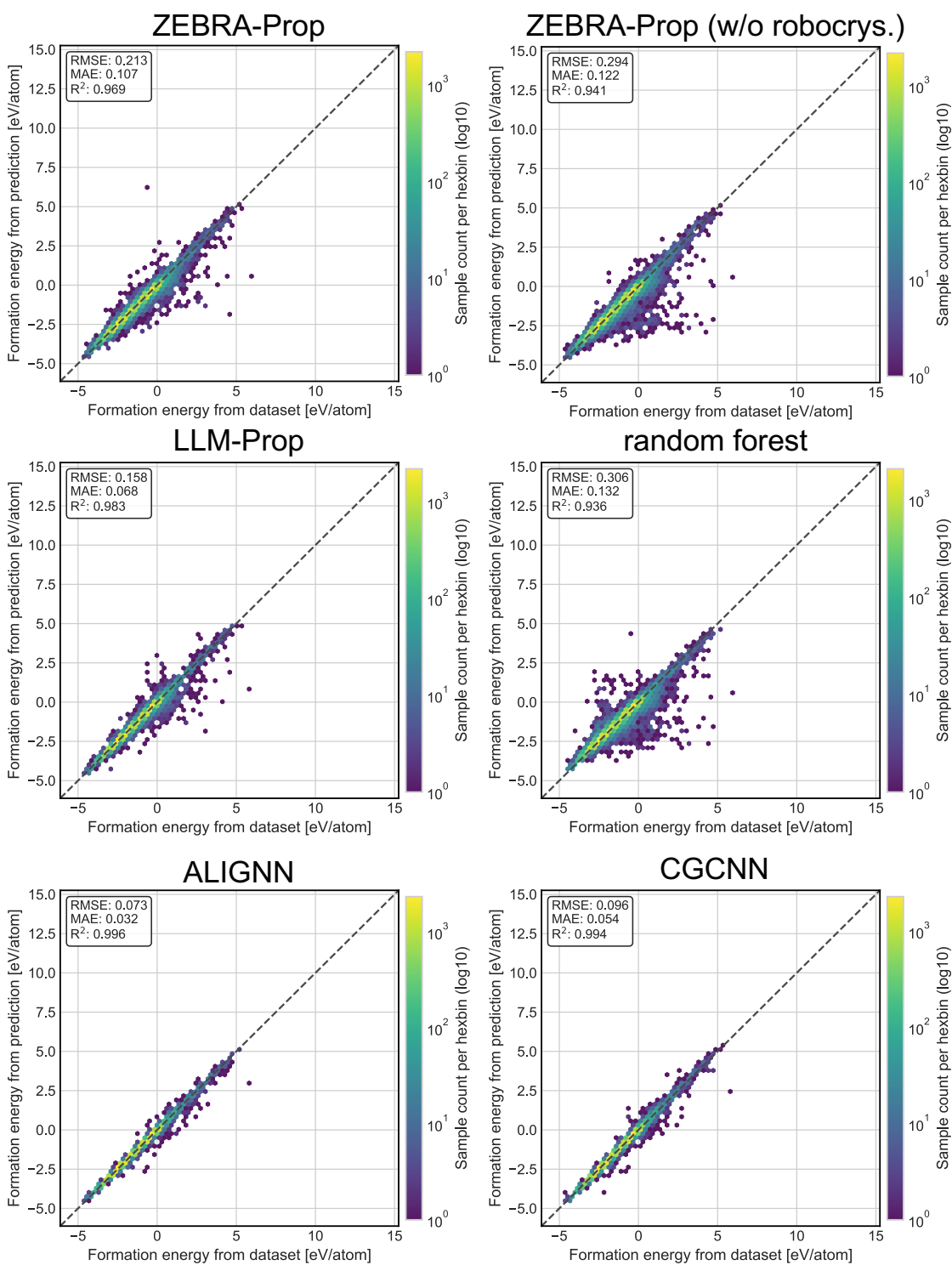

**Figure S3**(Continued).

## 5. Weights and Prediction Accuracy

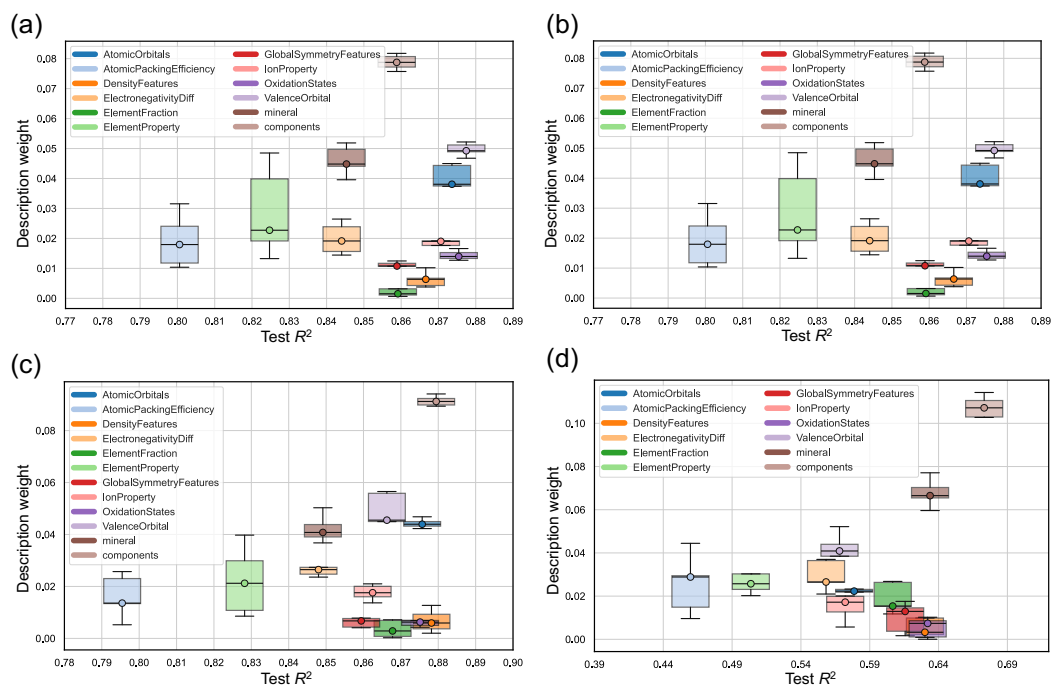

**Figure S4.** Relationship between single-sentence prediction performance ($R^2$ on the test fold) and the learned weights obtained from ZEBRA-Prop trained using all descriptions for the (a) band gap, (b) formation energy, (c) electronic contribution to the dielectric constant ($\log_{10}$ scale), and (d) ionic contribution to the dielectric constant ($\log_{10}$ scale) on the in-house dataset. Single-sentence prediction performance ($R^2$) is averaged over test folds in 5-fold cross-validation, while the learned weights are shown as box plots across folds.

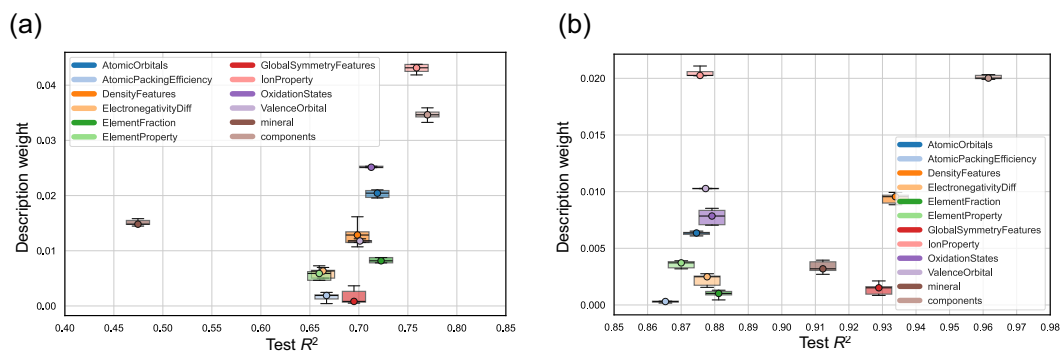

**Figure S5.** Relationship between single-sentence prediction performance ($R^2$ on the test fold) and the learned weights obtained from ZEBRA-Prop trained using all descriptions for the (a) band gap and (b) formation energy on the TextEdge dataset. Single-sentence prediction performance ($R^2$) is averaged over test folds in 5-fold cross-validation, while the learned weights are shown as box plots across folds.

## 6. Similarity of the Embeddings for Each Description

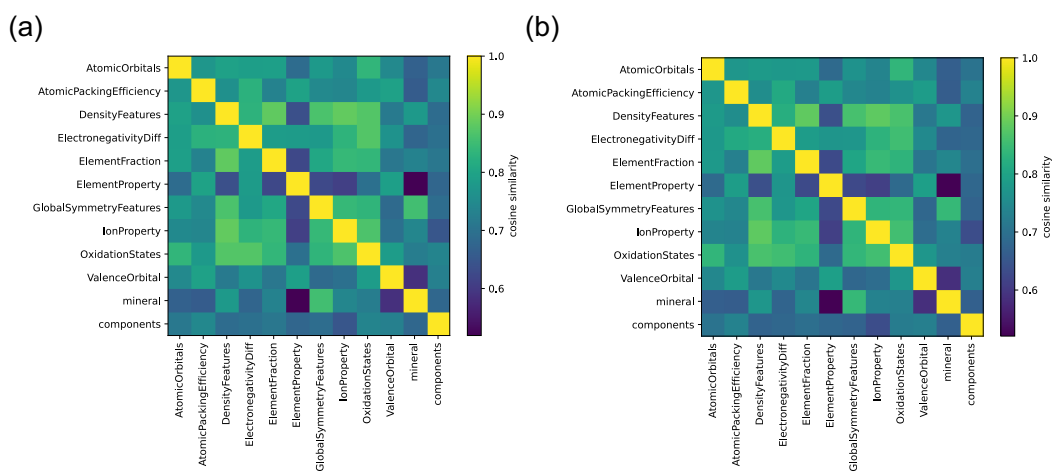

**Figure S6.** Average pairwise cosine similarity between input sentence embeddings. For each sample, cosine similarities are computed between all pairs of embeddings and then averaged over the entire dataset. Results are shown for the (a) in-house and (b) TextEdge datasets.

## 7. Weights and Prediction Accuracy

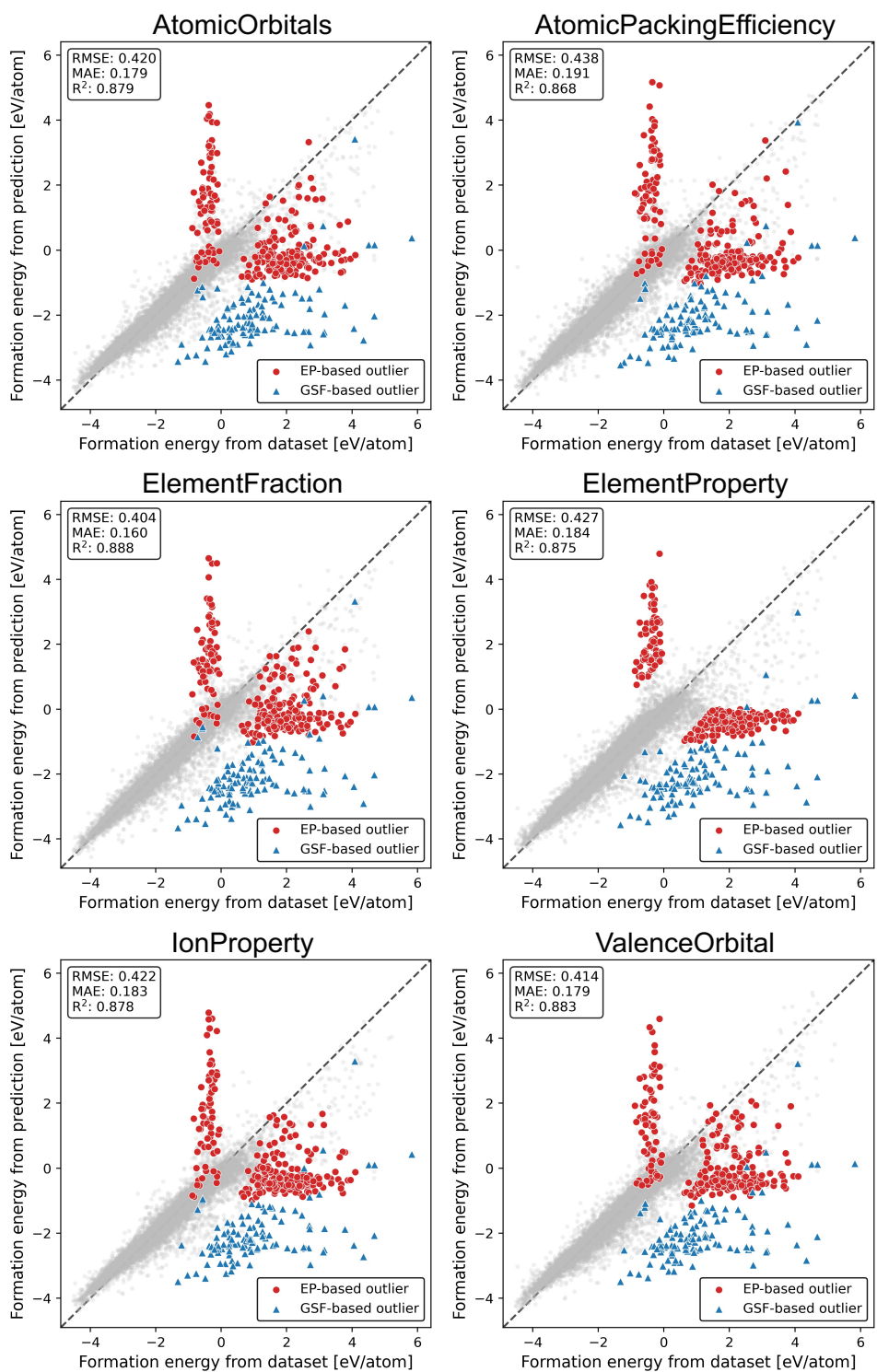

**Figure S7.** Formation energy prediction performance of ZEBRA-Prop using individual descriptions for the test set on the TextEdge dataset.

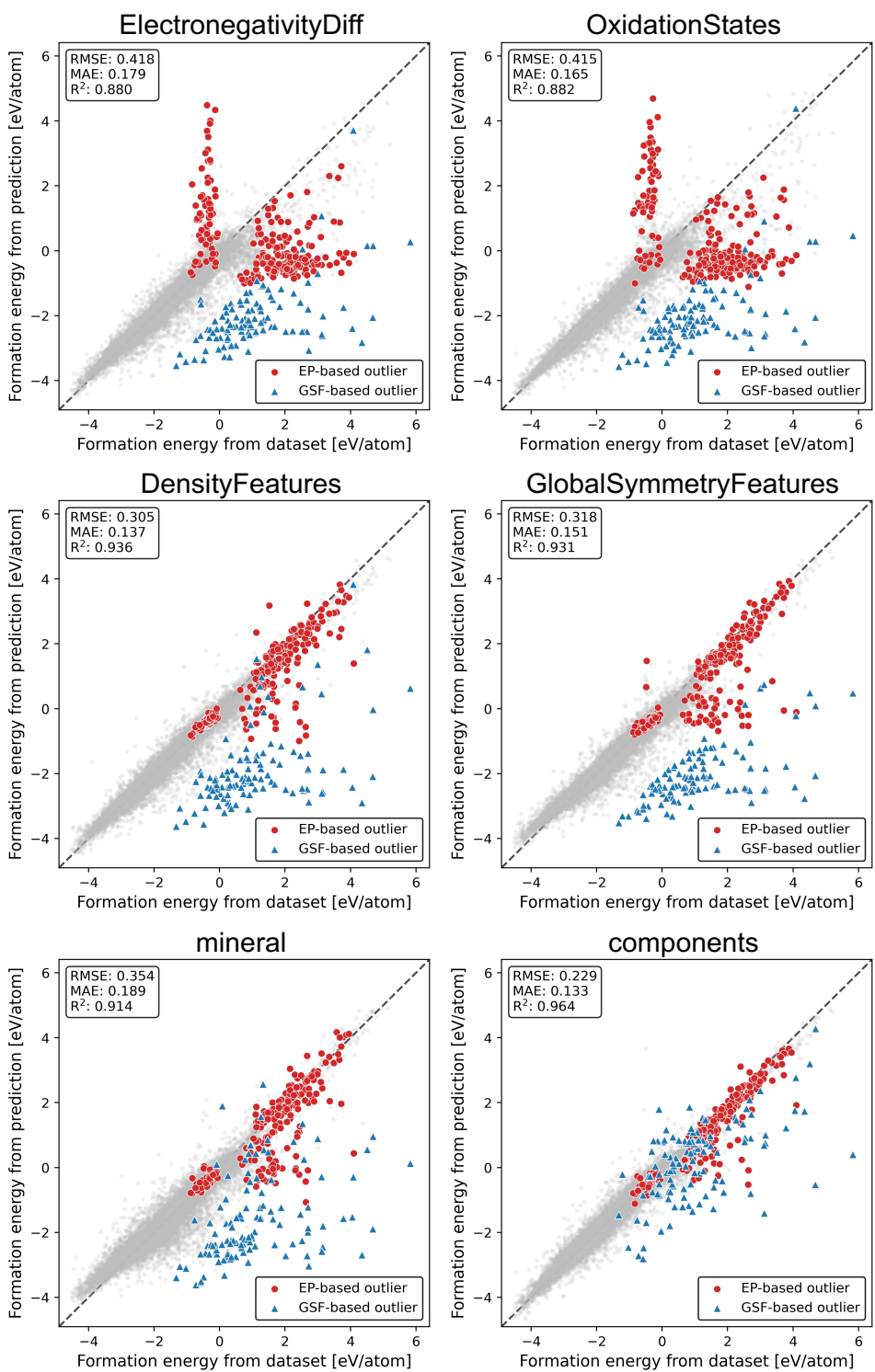

**Figure S7**(Continued).

## 8. Hyperparameter Grid Search

Based on the grid search results presented in **Figure S8-S12**, a learning rate of 0.001 and a batch size of 64 were selected, as they provided high predictive performance while maintaining computational efficiency.

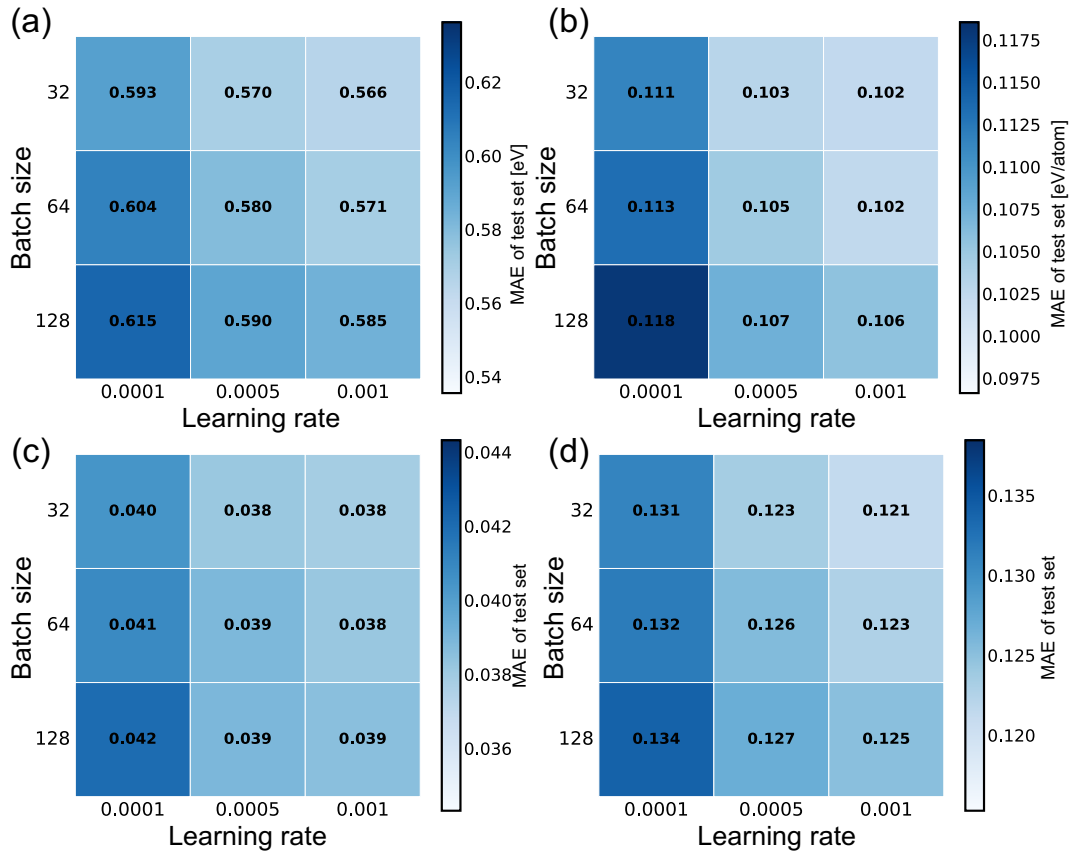

**Figure S8.** Performance obtained from a systematic hyperparameter optimization via grid search for ZEBRA-Prop on the in-house dataset: (a) the band gap, (b) the formation energy, (c) the electronic contribution to the dielectric constant (log10 scale), and (d) the ionic contribution to the dielectric constant (log10 scale). The reported values correspond to the mean test-set MAE across the test folds in 5-fold cross-validation.

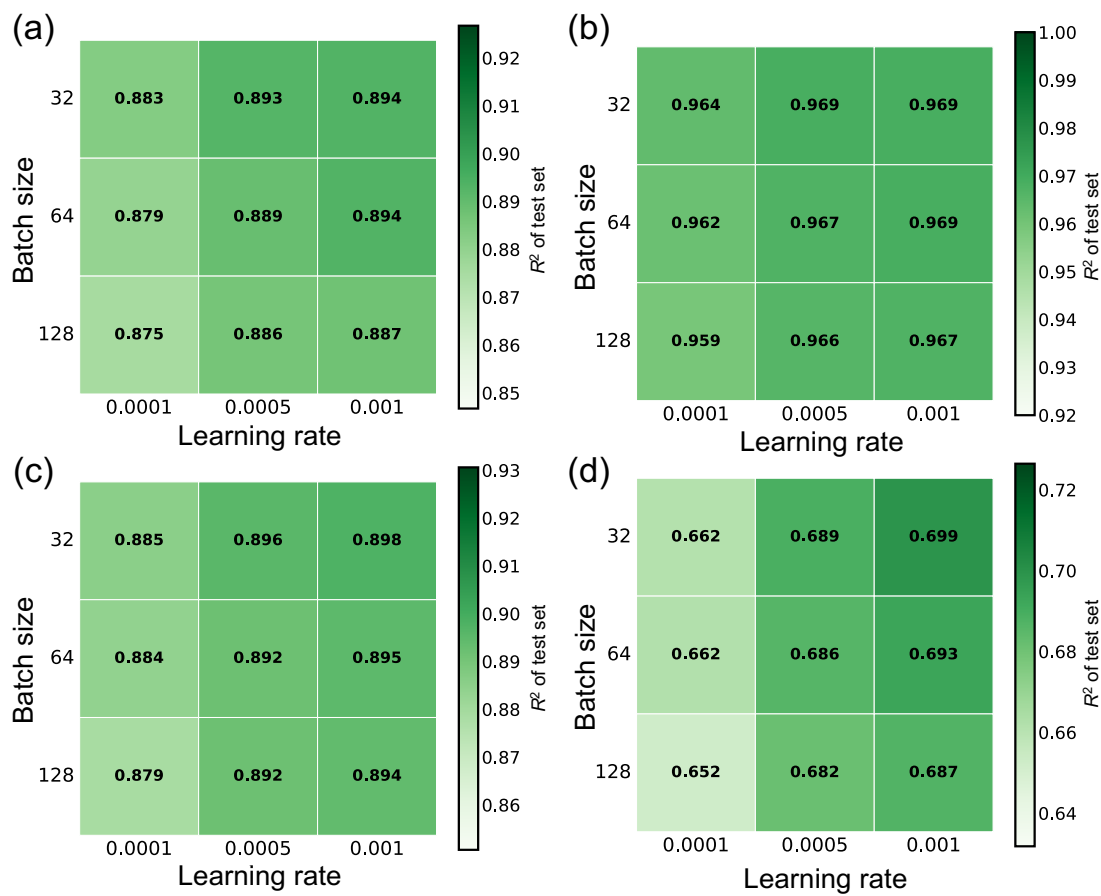

**Figure S9.** Performance obtained from a systematic hyperparameter optimization via grid search for ZEBRA-Prop on the in-house dataset: (a) the band gap, (b) the formation energy, (c) the electronic contribution to the dielectric constant (log10 scale), and (d) the ionic contribution to the dielectric constant (log10 scale). The reported values correspond to the mean test-set $R^2$ across the test folds in 5-fold cross-validation.

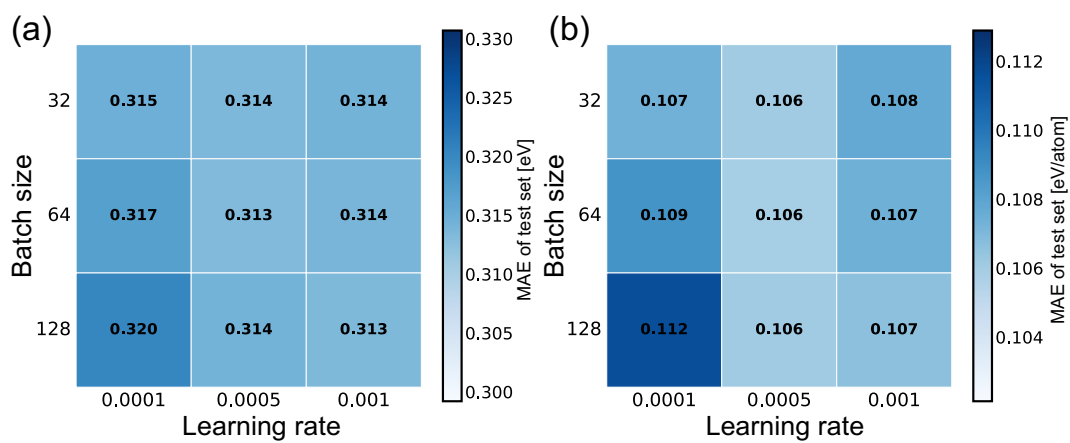

**Figure S10.** Performance obtained from a systematic hyperparameter optimization via grid search for ZEBRA-Prop on the in-house dataset: (a) the band gap and (b) the formation energy. The reported values correspond to the mean test-set MAE across the test folds in 5-fold cross-validation.

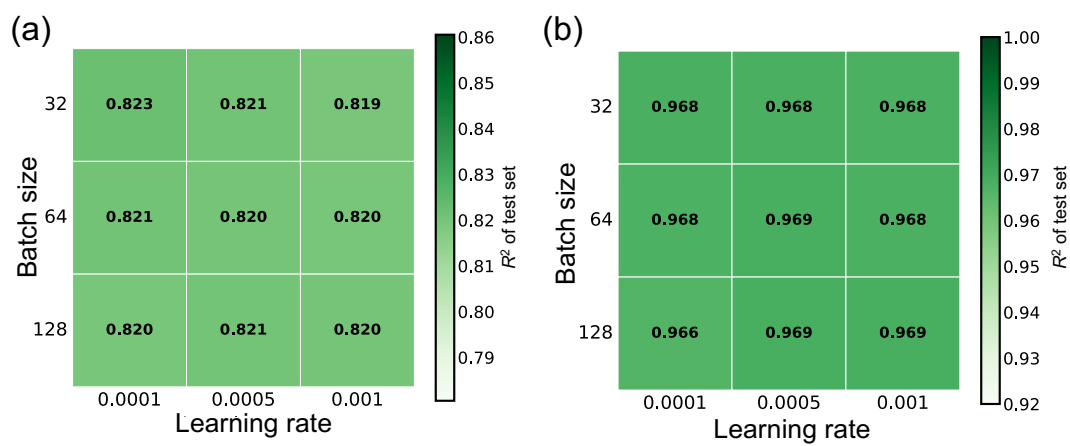

**Figure S11.** Performance obtained from a systematic hyperparameter optimization via grid search for ZEBRA-Prop on the in-house dataset: (a) the band gap and (b) the formation energy. The reported values correspond to the mean test-set $R^2$ across the test folds in 5-fold cross-validation.

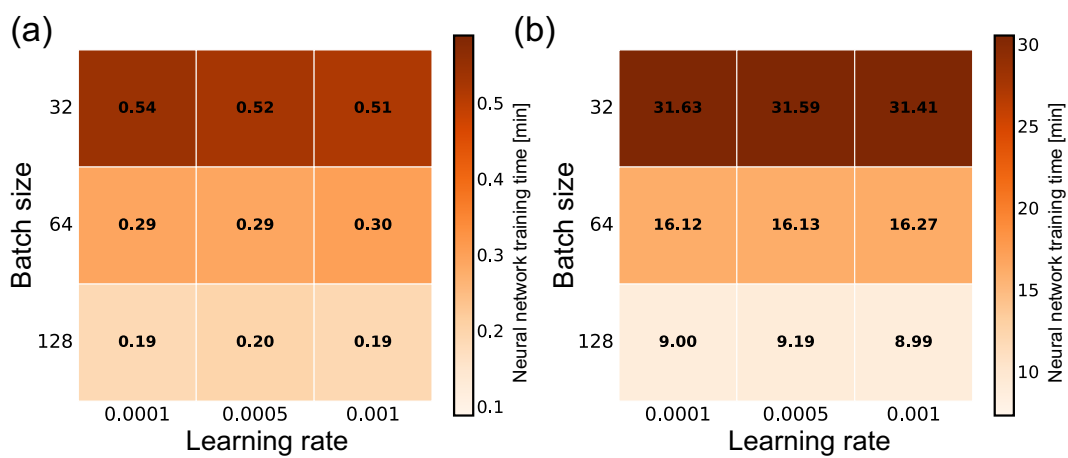

**Figure S12.** Performance obtained from a systematic hyperparameter optimization via grid search for ZEBRA-Prop on (a) the in-house and (b) TextEdge datasets. The reported values correspond to the mean neural network training time across the 5-fold cross-validation.